\documentclass[aps,pra,10pt,amsmath,eqsecnum,notitlepage,longbibliography,twocolumn,superscriptaddress]{revtex4-1}
\usepackage[colorlinks=true,urlcolor=NavyBlue,citecolor=NavyBlue,linkcolor=NavyBlue]{hyperref}
\usepackage[dvipsnames]{xcolor}
\usepackage{amssymb}
\usepackage{amsmath}
\usepackage{amsfonts}
\usepackage{graphicx}

\newcommand{\real}{\operatorname{Re}}
\newcommand{\imag}{\operatorname{Im}}
\newcommand{\parti}[2]{\frac{\partial #1}{\partial #2}}
\newcommand{\partit}[2]{\frac{\partial^2 #1}{\partial #2^2}}
\newcommand{\diff}[2]{\frac{d #1}{d #2}}

\newcommand{\ket}[1]{|#1\rangle}

\newcommand{\bra}[1]{\langle#1|}

\newcommand{\Avg}[1]{\left\langle#1\right\rangle}

\newcommand{\abs}[1]{\left|#1\right|}
\newcommand{\bk}[1]{\left(#1\right)}
\newcommand{\Bk}[1]{\left[#1\right]}
\newcommand{\BK}[1]{\left\{#1\right\}}

\newcommand{\trace}{\operatorname{tr}}

\newcommand{\expect}{\mathbb E}
\begin{document}
%\twocolumn[
\title{Testing quantum mechanics: a statistical approach}

\author{Mankei Tsang}

\email{eletmk@nus.edu.sg}
\affiliation{Department of Electrical and Computer Engineering,
  National University of Singapore, 4 Engineering Drive 3, Singapore
  117583}

\affiliation{Department of Physics, National University of Singapore,
  2 Science Drive 3, Singapore 117551}

\date{\today}

\begin{abstract}
  As experiments continue to push the quantum-classical boundary using
  increasingly complex dynamical systems, the interpretation of
  experimental data becomes more and more challenging: when the
  observations are noisy, indirect, and limited, how can we be sure
  that we are observing quantum behavior?  This tutorial highlights
  some of the difficulties in such experimental tests of quantum
  mechanics, using optomechanics as the central example, and discusses
  how the issues can be resolved using techniques from statistics and
  insights from quantum information theory.
\end{abstract}
%\pacs{42.50.Pq, 42.65.Ky, 42.65.Lm, 42.79.Hp}

\maketitle
\section{Introduction}
Once thought to be a theory confined to the atomic domain, quantum
mechanics is now being tested on increasingly macroscopic levels,
thanks to technological advances and the ingenuity of experimentalists
\cite{haroche_raimond,haroche_nobel,wineland_nobel,kippenberg,aspelmeyer,aspelmeyer2013}. As
experiments continue to push the quantum-classical boundary using
increasingly complex dynamical systems, the interpretation of
experimental data becomes more and more challenging: when the
observations are noisy, indirect, and limited, how can we be sure that
we are observing quantum behavior?
%This problem is complicated by the fact that quantum
%mechanics is often indistinguishable from classical stochastic
%models.
The goal of this tutorial is to highlight some of the difficulties in
such experimental tests of quantum mechanics and discuss how the
issues can be resolved using techniques from statistics and insights
from quantum information theory.  Apart from quantum physicists,
another target audience of this tutorial is statisticians and
engineers, who might be interested to learn more about quantum physics
and how statistics can be useful for the new generation of quantum
experiments.

The tutorial starts off in rather basic and general terms, introducing
the basic concepts of quantum mechanics in Sec.~\ref{qm} and
statistical hypothesis testing in Sec.~\ref{stat}. Sec.~\ref{versus}
is the centerpiece of this tutorial, discussing in detail why and how
quantum mechanics should be tested. To illustrate the concepts in the
context of recent experiments, optomechanics is used as the main
example.  Optomechanics refers to the physics of the interactions
between optical beams and mechanical moving objects. A moving mirror,
for example, will introduce varying phase shifts depending on its
position to an optical beam reflected by it.  The motion of the mirror
can then be inferred from measurements of the optical phase, while the
change in momentum of the reflected optical beam also means that the
mirror experiences a force, namely, radiation pressure. Optomechanics
technology has advanced so rapidly in recent years
\cite{kippenberg,aspelmeyer,aspelmeyer2013} that quantum effects are
becoming observable in mechanical devices with unprecedented sizes
\cite{brooks,purdy13,safavi2013b,purdy13b}. Such devices thus serve as
promising testbeds for new concepts in macroscopic quantum mechanics
\cite{chen2013}. Sec.~\ref{optomech} in particular studies the
optomechanics experiment reported by Safavi-Naeini \textit{et al.}\
\cite{safavi2012,safavi2013a} and demonstrates how statistics can be
applied to the experimental data.  For the motivated reader, the
Appendices also introduce some of the more advanced techniques in
classical and quantum probability theory that can facilitate the
experimental design and signal processing.

\section{\label{qm}Quantum mechanics}

\subsection{\label{quantal}Origin of quantum}
The word ``quantum'' in quantum mechanics refers to the fact that
certain physical quantities, such as energy and angular momentum,
exist only in discrete levels, or \emph{quanta}. This assumption,
together with classical mechanics, are able to explain many phenomena;
for example,
\begin{enumerate}
\item Planck's model of electromagnetic fields with discrete energy
  can explain the blackbody spectrum and, later by Einstein, the
  photoelectric effect.

\item Bohr's model of bound electrons with discrete energy and orbital
  angular momentum can explain the spectral lines of hydrogen.

\end{enumerate}
Despite its success, the seemingly ad-hoc nature of the quantal
assumption motivated theorists to find a deeper model. The result is
Schr\"odinger's wave mechanics and Heisenberg's matrix mechanics.

\subsection{The Hilbert-space theory}
The Schr\"odinger and Heisenberg pictures of quantum mechanics are
equivalent theories, which are able to explain the quantal model as a
consequence of deeper axioms based on Hilbert-space algebra. The
central quantities of the theory is the quantum state, which is a
complex vector denoted by $\ket{\psi}$, observables, which are
Hermitian matrices, and a unitary matrix $U$ for time evolution.

The Hilbert-space theory produces many predictions beyond the quantal
hypothesis. Perhaps the most outrageous one is the ``uncertainty''
relation, which states that the product of the variances of a pair of
incompatible observables, such as the position and momentum of an
electron, cannot be zero but is instead lower-bounded by a certain
positive value. The word ``uncertainty'' is put in quotes because, at
this stage, the ``uncertainty'' relation is nothing more than a
mathematical statement in the Hilbert-space theory. Although
Ehrenfest's correspondence principle tells us that the Hilbert-space
average of an observable obeys classical mechanics and gives us a
rough sense of how observables correspond to physical quantities, it
is unclear how the Hilbert-space variance is related to the common
sense of uncertainty, which is best described using probability
theory.

This problem becomes more apparent when one wishes to define the
correlation of incompatible observables. Correlation is a well defined
concept in probability, but in the Hilbert-space theory its definition
is ambiguous, with infinitely many ways of combining the observables
that result in different Hilbert-space moments.

An even more troubling problem with the theory is how to test it in an
experiment. In the Stern-Gerlach experiment, for example, an electron
beam interacts with magnetic fields, before being detected on a
screen. If we are to believe that the Hilbert-space theory is a
fundamental theory that governs all the interacting objects involved
in an experiment, then we must include in the Hilbert space not only
the electrons, but also the magnetic field, the screen, the
experimentalists themselves, and, by extension, the whole universe.

This viral nature of the Hilbert-space theory is nowadays taken more
seriously among some theorists.  On a pragmatic level, it makes the
theory, by itself, impossible to test experimentally, as the
experimentalists would have to take into account the universe,
including themselves, every time they would like to generate a
prediction from the Hilbert-space theory and perform an experiment to
test it.

To test the Hilbert-space theory, we must therefore find a way to
divorce the test object from the rest of the universe and extract
reproducible experimental results from the model. Fortunately, for
experimentalists, the von Neumann measurement theory provides a way
out.

\subsection{Quantum probability}
The von Neumann measurement theory provides a definition of quantum
measurement with respect to an observable, known as the von Neumann
measurement. The definition allows one to model a test object using a
Hilbert space, but still describe the rest of the universe as an
observer that follows the classical rules of probability. The
probabilities of measurement outcomes are determined from a
Hilbert-space model using Born's rule.
%For an observable with
%eigenbasis $\{\ket{Y}\}$, the Born probability for an outcome $Y$ is
%\begin{align}
%P(Y) &= \abs{\avg{Y|\psi}}^2.
%\end{align}
Although each measurement outcome is random, the Born probability
values are deterministic and can be estimated with increasing accuracy
by repeated experiments.  As the probabilities depend on the
Hilbert-space model being assumed, one can then obtain asymptotically
reproducible results that verify the validity of the Hilbert-space
theory. The combined theory of Hilbert space and von Neumann
measurement is referred to as the quantum probability theory.

With the quantum probability theory, the Hilbert-space moments and the
uncertainty relation acquire operational meanings: one can define
Hilbert-space averages in an unambiguous fashion by specifying the
measurements and asking how the averages are related to the expected
values for the measurements. Most importantly, the theory enables
experimentalists to stay safely in the realm of classical logic and
still test the Hilbert-space theory by considering smaller models.

We now have a quantum theory that predicts probabilities as verifiable
deterministic numbers, but it is very clumsy to use, as it provides no
rule that specifies which part of the experiment should be included in
the Hilbert space and which part should be defined as the observer.
This dichotomy is known as the Heisenberg cut. An empirical way of
deciding on a cut is as follows:
\begin{enumerate}
\item Make a guess of how the cut should be made and compute the
  quantum probabilities based on the cut.

\item The validity of the cut can be checked by making a larger cut:
  include more experimental objects in a larger Hilbert space, do the
  calculation again, and see if the predictions match.

\item Alternatively, one can also attempt to find smaller cuts with
  smaller Hilbert spaces (by using certain tricks known as the open
  quantum system theory).
\end{enumerate}
The arbitrariness of the cut is unsatisfactory to some, but we may
take a pragmatic view of the cutting procedure as an algorithm for the
scientific method. Without it, the very definition of scientific
experiments is endangered.

Much like the Hilbert-space theory superseding the quantal hypothesis,
there have been many proposals that claim to interpret or supersede
the quantum probability theory. Until such theories provide
distinguishable predictions, however, it is impossible to test them in
an experiment.

The concepts discussed thus far can be found in many standard
textbooks, for example, Ref.~\cite{peres}.
Appendices~\ref{qpt}--\ref{qest} present some of the more advanced
concepts and methods in quantum probability theory.

\section{\label{stat}Statistical hypothesis testing}
\subsection{Why bother?}
How do we test a hypothesis that gives only probabilities of the
measurement outcomes? An easy and by far the most common approach is
to perform an experiment in many trials or for a very long time, and
combine the outcomes into fewer numbers known as the test statistics,
such as the mean, correlation, or power spectral density.  The test
statistics are then compared with the expected values according to the
hypothesis.

To justify this averaging approach, one can appeal to the law of large
numbers or the ergodic theorem for the convergence of the test
statistics. Unfortunately, such laws are exact only for an infinite
number of trials or an infinitely long time.  These limits are called
``asymptotically almost surely'' in the lingo of probability theory,
but they also imply that, in finite time, we can never be sure, and a
way of characterizing the uncertainties is needed.

An analysis of experimental uncertainties is a standard prerequisite
for publication nowadays, but it is often treated more as an
afterthought than as an important part of research. Performing a
statistical analysis with utmost rigor is not only a moral
responsibility, but also has many benefits:
\begin{enumerate}
\item \textit{Experimental design}. Before implementing an experiment,
  it can tell experimentalists how much information they can gain from
  a setup, such that the design can be rejected, adopted, or improved,
  saving time, effort, and money.

\item \textit{Signal processing}. After the results are obtained,
  statistical signal processing techniques can be used to optimize
  their accuracy further and compute their errors.

\item \textit{Universality}. By using standard error measures, it is
  easier to compare and communicate the significance of an experiment.
  This is especially important for multidisplinary science and
  engineering applications.

\item \textit{Confidence}. Statistics can provide a measure of
  confidence, such that the experimentalists and the society in
  general can understand the value of the results and guard against
  the risks.

\item \textit{Fun}. Statistics is a full-fledged scientific discipline
  in itself, and many scientists and engineers find it fun to learn
  and apply.

\item \textit{Insight}. Learning about statistics may shed new light
  on the foundations of quantum probability theory.
\end{enumerate}
The last point should especially incentivize quantum physicists to
learn more about statistics.

\subsection{Bayesian hypothesis testing}
An intuitive approach to statistical hypothesis testing is known as
Bayesian hypothesis testing, which computes the posterior probability
$P(\mathcal H_j|Y)$ that a hypothesis $\mathcal H_j$ is
true given the observation $Y$ via the Bayes theorem:
\begin{align}
P(\mathcal H_j|Y) &= \frac{P(Y|\mathcal H_j)P(\mathcal H_j)}
{\sum_j P(Y|\mathcal H_j)P(\mathcal H_j)},
\label{bayes}
\end{align}
where $P(Y|\mathcal H_j)$ is the probability of the observation
predicted by a hypothesis $\mathcal H_j$ and $P(\mathcal H_j)$ is the
prior probability. A common criticism of the Bayesian method is that
the prior probabilities may imply subjective beliefs, but many
definitions of objective priors have been proposed and are now widely
accepted \cite{berger,bernardo,jaynes}. Some popular objective priors
are reviewed in Sec.~\ref{prior}.

If one is uncomfortable with priors, he can avoid them by turning to
frequentist methods. The significance of a frequentist test is much
more difficult to comprehend and communicate to others, however,
unlike the much more intuitive meaning of a posterior probability. For
example, a popular frequentist significance measure is called the
$p$-value, which is the probability that a test statistic would be
more extreme than the experimentally obtained value if a hypothesis to
be rejected is true.

At least one alternative is needed to compute the posterior
probability distribution. If there is no obvious alternative and one
lacks the imagination to come up with one, it is possible to compare a
hypothesis with reference alternatives based on more mathematical
grounds \cite{jaynes}. Fortunately, for quantum tests, alternatives,
such as classical mechanics and hidden-variable models, are abundant.

% In practice, even the most pedantic frequentists would agree that
% Bayesian methods are a significant improvement over the current
% standard of error analysis.
The rest of the tutorial will focus on the Bayesian theory. For
critiques of frequentist methods, see
Refs.~\cite{berger,bernardo,jaynes}.

\subsection{\label{strength}Strength of an experiment}
To judge the significance and value of an experiment, it is useful to
quantify how strongly an experimental result may sway one's opinion.
For the simplest example, consider two hypotheses.  The ratio of
the posterior probabilities is
\begin{align}
\frac{P(\mathcal H_1|Y)}{P(\mathcal H_0|Y)} &= 
\frac{P(Y|\mathcal H_1)P(\mathcal H_1)}{P(Y|\mathcal H_0)P(\mathcal H_0)} = 
\Lambda(Y)\frac{P(\mathcal H_1)}{P(\mathcal H_0)},
\\
\Lambda(Y) &\doteq 
\frac{P(Y|\mathcal H_1)}{P(Y|\mathcal H_0)}.
\end{align}
$\Lambda(Y)$ is called the likelihood ratio. It is used to update
one's prior beliefs about the two hypotheses, and can be understood as
the strength of a given evidence $Y$ for one hypothesis against the
other. An experiment shows strong evidence for $\mathcal H_1$ against
$\mathcal H_0$ when $\Lambda(Y) \gg 1$ and vice versa when $\Lambda(Y)
\ll 1$.
%The likelihood ratio is also useful for frequentist tests,
%which in general involve comparing the ratio against a threshold.

Unless the two hypotheses predict the same probability distribution,
the likelihood ratio cannot be computed until some results are
obtained. For experimental design, it is useful to know in advance how
much the likelihood ratio is expected to rise or fall. One measure
that quantifies this expected information is the relative entropy:
\begin{align}
D(P_1||P_0) &\doteq \expect\Bk{\ln\Lambda(Y)|\mathcal H_1}
\\
&=
\sum_Y P(Y|\mathcal H_1) \ln \frac{P(Y|\mathcal H_1)}{P(Y|\mathcal H_0)},
\end{align}
where $\expect$ denotes the expected value.  To see why it is a
sensible measure of information, consider $M$ independent trials with
observations $Y_1,Y_2,\dots,Y_M$, each generating a likelihood ratio
$\Lambda(Y_m)$.  The collective log-likelihood ratio is
\begin{align}
\ln\Lambda(Y_1,\dots,Y_M) &= \sum_{m=1}^M \ln \Lambda(Y_m).
\end{align}
This means that, as $M\to\infty$, if the trials have identical
probability distributions and $\mathcal H_1$ is true,
\begin{align}
\ln\Lambda(Y_1,\dots,Y_M) &\to M \expect\Bk{\ln \Lambda(Y_m)|\mathcal H_1}
= MD(P_1||P_0).
\end{align}
Since the relative entropy is always nonnegative, the ratio is
expected to rise if $\mathcal H_1$ is true.  The same argument works
also if $\mathcal H_0$ is true and the log-likelihood ratio should
fall, since $-M D(P_0||P_1) \le 0$.  The relative entropies thus
provide the experimentalist an idea of how the expected strength of an
experiment increases with the number of trials. This rise of expected
information is important, as it tells us that, even if each trial is
uncertain, more evidence will get us closer to the truth.

For other operational meanings of the relative entropy, see
Refs.~\cite{bernardo,cover}. For multiple-hypothesis testing in
general, an appealing measure of information gain is the mutual
information; see Ref.~\cite{bernardo}.

\subsection{\label{decision}Making decisions}
For engineering applications, including communication, robotic
control, and financial trading, the goal of hypothesis testing is not
only to gain knowledge or convince skeptics, but also to make a
decision on one hypothesis. We define a decision rule as $\mathcal
H_k(Y)$ and the penalty or cost incurred by a decision on $\mathcal
H_k$ when $\mathcal H_j$ is true via the loss function $L(\mathcal
H_j,\mathcal H_k)$.  The expected loss is called the risk of a
decision rule \cite{berger}:
\begin{align}
\mathcal R(\mathcal H_j) &\doteq 
\sum_Y L(\mathcal H_j,\mathcal H_k(Y))P(Y|\mathcal H_j).
\end{align}
If we average the risk function over a prior, we obtain the so-called
Bayes risk:
\begin{align}
  \mathcal R &\doteq \sum_j \mathcal R(\mathcal H_j)P(\mathcal H_j),
\label{R}
\end{align}
which can also be written in terms of the posterior distribution as
\begin{align}
\mathcal R &= \sum_{Y}\Bk{\sum_j L(\mathcal H_j,\mathcal H_k(Y))
P(\mathcal H_j|Y)P(Y)}.
\label{R2}
\end{align}
To minimize $\mathcal R$, we can choose a $\mathcal H_k(Y)$ that
minimizes each of the square-bracketed terms in Eq.~(\ref{R2}).  This
is equivalent to a decision rule that minimizes the posterior expected
loss:
\begin{align}
\check{\mathcal H}(Y) &=
\arg \min_{\mathcal H_k(Y)}\sum_j L(\mathcal H_j,\mathcal H_k(Y))
P(\mathcal H_j|Y).
\label{bayes_decision}
\end{align}
This risk minimization serves as another motivation for the Bayesian
approach. For example, the probability of making a wrong decision
$P_e$, or the error probability for short, is equivalent to defining
the loss function as
\begin{align}
L(\mathcal H_j,\mathcal H_k) &= 1-\delta_{jk},
\end{align}
and the optimal decision is to choose the hypothesis with the highest
posterior probability $P(\mathcal H_j|Y)$.

Except for a few special cases, the error probability is hard to
compute exactly, but it can be sandwiched between a lower bound and an
upper bound in the case of two hypotheses.  For $P(\mathcal H_0) =
P(\mathcal H_1) = 1/2$, the bounds are given by
\cite{kailath69,vantrees}
\begin{align}
\frac{1}{2} \BK{1-\sqrt{1-\exp[-2C(0.5)]}}
\le\min_{\mathcal H_k(Y)}P_e 
\nonumber\\
\le \frac{1}{2}\min_{0\le s\le 1} \exp\Bk{-C(s)},
\label{sandwich}
\end{align}
where $C(s)$ is known as the Chernoff information:
\begin{align}
C(s) &\doteq -\ln \expect\Bk{\Lambda^s(Y)|\mathcal H_0},
\label{chernoff}
\end{align}
and $C(0.5)$ is called the Bhattacharyya distance. The Chernoff upper
bound is useful for guaranteeing the testing accuracy, while the lower
bound is more useful as a no-go theorem.  The Chernoff information can
be used to lower-bound the relative entropy as well:
\begin{align}
D(P_1||P_0) &\ge \max_{0\le s \le 1}\frac{C(s)}{1-s}.
\end{align}
Due to its decision-theoretic meaning for a finite number of trials
and the asymptotic tightness of the upper bound in
Eq.~(\ref{sandwich}) \cite{levy}, the Chernoff information is
considered a more meaningful information measure than the relative
entropy, although the former is often more difficult to compute.

% Another important application of decision theory is the theory of
% games, where players bet on different outcomes and formulate
% strategies based on how loss is incurred and the players' opinions
% about the risks. Although games, and gambling in particular, are one
% of the original motivations of Bayesian methods, Bayesian theory has
% since outgrown the narrow application and can be formulated on more
% fundamental logical principles \cite{jaynes}.

For more details about decision theory, see Ref.~\cite{berger}.  For a
discussion of decision theory in the context of scientific methods,
see Ref.~\cite{jaynes}. Shannon information theory should really be
called communication theory and may be regarded as a branch of
decision theory; see Ref.~\cite{cover}. For the use of decision theory
for engineering applications, see Ref.~\cite{vantrees,levy}.

\subsection{\label{parameter}Parameter estimation}
Instead of considering just two hypotheses, let us consider the other
extreme, where a continuum of hypotheses may be assumed, and rewrite
the assumptions as a column vector of parameters $\theta$. The problem
then becomes a parameter estimation problem.  $\theta$ can be
estimated by computing the posterior probability density
$P(\theta|Y)$:
\begin{align}
P(\theta|Y) &= \frac{P(Y|\theta)P(\theta)}
{\int d\theta P(Y|\theta)P(\theta)},
\end{align}
where $P(\theta)$ is the prior probability density.  As a measure of
posterior uncertainty, a \emph{credible region} for $\theta$ can be
defined as the set $\Theta_c(Y)$ with a high posterior probability
$P_c$ \cite{bernardo}:
\begin{align}
\int_{\theta\in\Theta_c}d\theta P(\theta|Y)  &= P_c,
\label{credible}
\end{align}
say, $95\%$.  This allows us to dismiss the region outside $\Theta_c$
as improbable. Another common measure useful for defining error bars
is the posterior mean and covariance matrix:
\begin{align}
\check\theta(Y) &= \expect(\theta|Y)
= \int d\theta \theta P(\theta|Y),
\\
\Pi(Y) &=
\expect\Bk{(\theta-\check\theta)(\theta-\check\theta)^\top|Y}
\nonumber\\
&=\int d\theta (\theta-\check\theta)(\theta-\check\theta)^\top P(\theta|Y),
\end{align}
where $^\top$ denotes the matrix transpose. 

A decision rule, called an estimator in this context, can also be
obtained by specifying a loss function. For example, the mean-square
error matrix is
\begin{align}
\Sigma &\doteq \expect\Bk{(\theta-\check\theta)(\theta-\check\theta)^\top}
\nonumber\\
&=   \int d\theta \sum_Y
(\theta-\check\theta)(\theta-\check\theta)^\top P(Y|\theta)P(\theta),
\end{align}
which is minimized if we decide on the posterior mean.  Like the error
probability $P_e$, $\Sigma$ is usually difficult to compute exactly,
so one often has to resort to approximations or bounds. The most
popular information measure for parameter estimation is the Fisher
information matrix $J(\theta)$, defined as
\begin{align}
J_{jk}(\theta) &\doteq \sum_Y P(Y|\theta) 
\Bk{\parti{}{\theta_j}\ln P(Y|\theta)}
\Bk{\parti{}{\theta_k}\ln P(Y|\theta)}.
\label{fisher}
\end{align}
A useful identity is \cite{vantrees3}
\begin{align}
J_{jk}(\theta) &= 4 \Bk{\parti{^2}{\theta_j\partial\theta_k}
C(0.5,\theta,\theta')}_{\theta'=\theta},
\label{JC}
\end{align}
where $C(0.5,\theta,\theta')$ is the Bhattacharyya distance given by
Eq.~(\ref{chernoff}) with $P(Y|\mathcal H_0) = P(Y|\theta)$ and
$P(Y|\mathcal H_1) = P(Y|\theta')$.  The Fisher information determines
general lower limits on the mean-square errors via the Cram\'er-Rao
family of bounds \cite{vantrees,bell}.  The Bayesian version is given
by the following matrix inequality:
\begin{align}
\Sigma &\ge \bk{J + J_{\rm prior}}^{-1},
\\
J_{\rm prior} &\doteq \int d\theta P(\theta)\Bk{\parti{}{\theta_j}\ln P(\theta)}
\Bk{\parti{}{\theta_k}\ln P(\theta)},
\\
J &\doteq \int d\theta P(\theta) J(\theta),
\end{align}
and is valid for any estimator. $J$ can then give us an idea of how
accurate an experiment can be in resolving the parameters. An
alternative family of lower bounds called the Ziv-Zakai bounds can
also be computed using $C(0.5,\theta,\theta')$ and are often tighter
than the Cram\'er-Rao bounds \cite{bell,qzzb}.

\subsection{\label{prior}Objective priors}
For scientific tests, it is preferable to choose a prior distribution
based on objective principles. One such principle is maximum entropy
\cite{jaynes}, which chooses the prior that maximizes the entropy
$-\sum_jP(H_j)\ln P(H_j)$ in the presence of known constraints about
$P(H_j)$. Justifications of this approach can be found in
Refs.~\cite{jaynes,shore}.  For parameter estimation, a more popular
choice is the Jeffreys prior \cite{berger,bernardo}:
\begin{align}
P(\theta) &\propto \sqrt{\det J(\theta)},
\label{jeffreys}
\end{align}
where $J(\theta)$ is the Fisher information matrix given by
Eqs.~(\ref{fisher}). It has the advantage of giving the same
probability measure $P(\theta)d\theta$ regardless of how the unknown
parameters are defined.

One may also resort to decision theory and choose the so-called least
favorable prior, which maximizes the Bayes risk given by Eq.~(\ref{R})
for the Bayes decision rule given by Eq.~(\ref{bayes_decision})
\cite{berger}. It is the most conservative prior in the context of
decision theory and has the advantage of producing a Bayes decision
rule that coincides with the frequentist minimax rule \cite{berger},
but it is often much more difficult to calculate than the other
priors.

For more in-depth discussions of objective priors, see
Refs.~\cite{berger,bernardo,jaynes}.

\section{\label{versus}Quantum versus classical}
\subsection{Classical mechanics}
Classical mechanics is a natural alternative hypothesis for quantum
tests. Experiments and observations have verified its validity on a
macroscopic level, such that one should assign a significant value for
its prior probability. This prior cannot be too high either, as the
quantum theory has also been well tested for simple systems, and many
theorems rule out naive classical mechanics if the quantum theory is
true. A ``quantum versus classical'' test is thus most interesting on
a complexity level where the prior probabilities are comparable, if
not equal.

Even if one does not personally believe in one of the theories on the
level being tested, there are many reasons why the verification of a
particular hypothesis is relevant to science and engineering:
\begin{enumerate}
\item \emph{Learning curve}. Many people understand classical
  mechanics but quantum mechanics takes much more effort to learn. If
  classical mechanics is sufficient, the quantum model would be
  unnecessary for them.
  % This attitude is especially prevalent in engineering, as evidenced
  % by the many textbooks (for example,
  % Refs.~\cite{goodman_stat,verdeyen,yariv_yeh,streetman}) that try
  % to avoid the Hilbert-space theory.

\item \emph{Quantum simulation}. Even if one knows quantum mechanics,
  solving it for macroscopic objects is still very hard. With current
  computers, classical mechanics can take much less resources to solve
  than known numerical methods for quantum mechanics.

\item \emph{Quantum computing}. For a few problems, such as factoring
  large numbers, it has been suggested that a quantum computer can be
  superior to a classical one \cite{nielsen}. Quantum simulations
  might also be easier on a quantum computer. A test of quantum
  mechanics on a macroscopic level would shed light on the feasibility
  of a practical quantum computer.

\item \emph{Quantum information}. Many limits on sensing and
  communication have been derived based on the quantum probability
  theory
  \cite{helstrom,holevo,braginsky,caves,wiseman_milburn,nielsen,paris_rehacek,glm_science,glm2011,twc,stellar,qzzb,tsang_nair,tsang_open},
  whereas classical mechanics is fundamentally deterministic. Emergent
  determinism would be good news for sensing near the quantum limits
  but bad news for quantum security protocols.

\item \emph{Quantum gravity}. There are alternative theories about how
  gravity might modify quantum mechanics on a macroscopic level
  \cite{pikovski,chen2013,blencowe}. Such theories may be modeled
  using classical mechanics.
\end{enumerate}
To clarify these issues, we should search for a classical mechanics
model that is as close to the quantum theory as possible, such that,
without an experiment, one has no evidence for one over the other, and
the experiment can provide new and useful information that people do
not already know.

To find ``the most quantum'' classical model, the correspondence
principle is helpful in the first order, but becomes ambiguous when
one attempts to relate higher-order Hilbert-space moments to classical
statistics. To prevent prior intuition from limiting our imagination
and cast a wider net, it is sometimes worthwhile to adopt a more
abstract mathematical approach. The theory of quantum computation
turns out to be useful in this way.

\subsection{Classical simulability}
One of the most general results about equivalent models from the
quantum and classical theories is the Gottesman-Knill theorem
\cite{nielsen} and its generalizations for continuous variables
\cite{braunstein_rmp,bartlett2012}.  The rough idea is that a certain
class of models under the quantum probability theory is equivalent to
classical hidden Markov models (HMM) \cite{elliott}, with restrictions
on the number of dimensions of the classical state space and the
number of time steps. ``Restrictions'' is the key word here, as even
the full quantum probability model can in principle be simulated on a
classical computer, if one simply takes all the parameters that
specify the quantum model and use brute-force finite-element methods.

The classical simulability theorems are useful as no-go theorems: they
rule out the necessity of the full quantum theory when the system can
be described by a more succinct classical model.
% Such a classically simulable model can be surprisingly powerful and
% explain phenomena, such as the blackbody spectrum, that are thought
% to necessitate a quantal hypothesis \cite{boyer}.
The hidden variables in such a model can correspond to incompatible
observables; they obey uncertainty and measurement-disturbance
relations via additional constraints on the probability distributions
and system/observation noise sources.
%making the model a much larger class than the
%quantum-mechanics-free model discussed by Tsang and Caves \cite{qmfs}
%(see also earlier work by Koopman \cite{koopman,peres} and Gough and
%James \cite{gough}).

The HMM is invaluable for classical estimation and control
applications \cite{elliott} and provides the proper foundation for any
quantum versus classical debate. It is briefly reviewed in
Appendices~\ref{hmm}--\ref{hgmm}, which also set the stage for the
quantum probability theory that follows in
Appendix~\ref{qpt}--\ref{qest}.

\subsection{\label{uncertain}Testing the uncertainty principle}
Even for classically simulable systems, there are interesting quantum
features to be tested. A test showing a modification of the
uncertainty principle, for example, would be highly valuable to
quantum gravity theory and relevant to quantum sensing applications,
not to mention the Nobel prizes that are sure to follow, if the test
is done with rigor and accuracy and leads to new physics.

Let us therefore focus on a classically simulable system in this
section and use the HMM for all the hypotheses to be tested. Let $X$
be the hidden variables, and let's introduce additional parameters
$\theta$ that define the HMM as follows:
\begin{align}
  P(Y|\mathcal H_j) &= \int d\theta P(Y|\theta) P(\theta|\mathcal H_j),
\label{hidden_theta}\\
P(Y|\theta) &= \int dX P(Y,X|\theta).
\label{hidden_x}
\end{align}
For an optomechanics experiment for instance, $X$ can include the
canonical positions and momenta of optical and mechanical oscillators,
while $\theta$ can include the resonance frequencies, the damping
rates, the initial covariance matrix, and the system and observation
noise power levels. This breaking down of a hypothesis into a
hierarchy of more refined ones is very convenient for modeling and
numerical analysis in practice. $\mathcal H_j$ is then called a
composite hypothesis.

For now, the hypotheses $\mathcal H_j$ are assumed to differ only in
their prior assumptions about $\theta$ according to $P(\theta|\mathcal
H_j)$.  The quantum theory, for example, would manifest itself as
inequalities that imposes constraints on the allowable values of
$\theta$, while alternative quantum gravity theories
\cite{pikovski,chen2013,blencowe} may impose different
constraints. Such hard constraints can be imposed by specifying a
zero-probability set $\bar\Theta_j$:
\begin{align}
P(\theta|\mathcal H_j) = 0 \textrm{ for }\theta \in \bar\Theta_j.
\end{align}
Other prior information, such as independent calibrations, can also be
incorporated into $P(\theta|\mathcal H_j)$.

A constructive strategy for composite hypothesis testing is as follows:
\begin{enumerate}
\item Compute $P(Y|\theta)$ for all plausible $\theta$, taking any
  advantage offered by the hidden structure in Eq.~(\ref{hidden_x}).

\item Combine $P(Y|\theta)$ into $P(Y|\mathcal H_j)$ for each
  hypothesis, using the prior $P(\theta|\mathcal H_j)$ and
  Eq.~(\ref{hidden_theta}).

\item Compute the posterior probabilities $P(\mathcal H_j|Y)$ using
  the Bayes theorem given by Eq.~(\ref{bayes}).

\item $P(Y|\theta)$ can also be used for parameter estimation without
  assuming any composite hypothesis.
\end{enumerate}
A tutorial example of this Bayesian approach for optomechanics shall
be presented in the next section.

If one is uncomfortable with any choice of prior, $P(Y|\theta)$ can
also be used in frequentist tests. One example is the generalized
likelihood-ratio test \cite{levy}, which uses constrained
maximum-likelihood estimates of $\theta$ in $P(Y|\theta)$ instead of
the averaging.

\subsection{\label{optomech}An optomechanics example}
\subsubsection{Modeling}
Consider the experiment on a cavity optomechanical system by
Safavi-Naeini \textit{et al.}\ \cite{safavi2012,safavi2013a}. Let
$\omega_a$ be the resonance frequency of an optical cavity mode and
$\omega_b$ be that of a mechanical oscillator.  A continuous-wave
laser pump beam with detuned frequency $\omega_a-\omega_b$ is coupled
into the system, causing a parametric interaction between the optical
mode and the mechanical mode. The output optical field is then
measured via heterodyne detection. The goal of the experiment is to
infer properties of the mechanical oscillator motion from the noisy
optical measurements.

Define $a(t)$ as the complex analytic signal of the optical mode field
and $b(t)$ as that of the mechanical mode.  By considering the Wigner
representation of the output field, making appropriate rotating-wave
approximations, and adding excess output noise for the heterodyne
detection, we can obtain the following classical linear equations of
motion:
\begin{align}
\diff{a(t)}{t} &= igb(t) - \frac{\gamma_a}{2} a(t) + \sqrt{\gamma_a} A(t),
\label{aminus}
\\
\diff{b(t)}{t} &= ig^*a(t) - \frac{\gamma_b}{2} b(t) + \sqrt{\gamma_b} B(t),
\label{bminus}
\\
A_{-}(t) &= \sqrt{\gamma_a} a(t) - A(t) + A'(t),
\label{Aminus}
\end{align}
where $g$ is an optomechanical coupling constant proportional to the
field of the pump beam, $\gamma_a$ and $\gamma_b$ are the damping
rates of the optical and mechanical modes, respectively, $A(t)$ is an
optical input noise source, $B(t)$ is a mechanical noise source,
$A_{-}(t)$ is the output field near $\omega_a$ to be measured by
heterodyne detection, and $A'(t)$ is the excess output noise.  These
equations suggest that there is a coherent energy exchange between the
optical and mechanical modes enabled by the pump.

The noise sources are assumed to be zero-mean, phase-insensitive, and
uncorrelated with one another, with power levels defined by
\begin{align}
\expect\Bk{A(t)A^*(t')|\theta} &= S_A\delta(t-t'),
\\
\expect\Bk{B(t)B^*(t')|\theta} &= S_B\delta(t-t'),
\\
\expect\Bk{A'(t)A'^*(t')|\theta} &= S_A'\delta(t-t').
\end{align}
Steady-state initial conditions can also be assumed.  The derivation
of these classical equations of motion is a standard exercise in
quantum optics \cite{walls_milburn}; similar derivations have been
reported in
Refs.~\cite{khalili2012,jayich,safavi2013a,classical_model}. As
discussed in Appendices~\ref{hgmm} and \ref{wigner}, this model is
equivalent to a continuous-time hidden Gauss-Markov model (HGMM)
\cite{elliott}.

In another set of measurements, a blue-detuned pump beam with
frequency $\omega_a+\omega_b$ is used instead. The equations of
motion are
\begin{align}
\diff{a(t)}{t} &= igb^*(t) - \frac{\gamma_a}{2} a(t) + \sqrt{\gamma_a} A(t),
\label{aplus}
\\
\diff{b(t)}{t} &= iga^*(t) - \frac{\gamma_b}{2} b(t) + \sqrt{\gamma_b} B(t),
\label{bplus}
\\
A_{+}(t) &= \sqrt{\gamma_a} a(t) - A(t) + A'(t).
\label{Aplus}
\end{align}
These equations suggest a two-mode parametric amplification mechanism
that is different from the first experiment.  Note that this
hidden-variable model is similar to that for the first set of
measurements. This is a result of using the Wigner representation. If
the Sudarshan-Glauber or Husimi representations \cite{walls_milburn}
had been used instead, the model would have to be modified more
substantially, leading to needless complexity. The Wigner
representation can be used with minimal changes for homodyne detection
as well, so it is the best method at our disposal for deriving
classical dynamical models with the least amount of contextuality; see
Appendices~\ref{dwigner} and \ref{wigner} for further discussions
about the Wigner representations.

For simplicity, we assume that the parameters have not drifted from
those in the first set of measurements, and $g$, $\gamma_a$,
$\gamma_b$, and $S_A'$ are so accurately determined prior to the
experiments that they can be regarded as being known exactly. The only
unknowns included in $\theta$ are the system noise power levels:
\begin{align}
\theta &= \bk{\begin{array}{c}S_A\\ S_B\end{array}},
\end{align}
and we seek to perform hypothesis testing and parameter estimation
based on the information gained about these parameters from the
measurements.

\subsubsection{Power spectral densities}
Before discussing the statistical hypothesis testing method, let us
first consider the expected infinite-time statistics. The most
important ones are the power spectral densities:
\begin{align}
S_\pm(\omega|\theta) &\doteq \lim_{T\to\infty}
\expect\Bk{\frac{1}{T}\abs{\int_{0}^{T} dt A_\pm(t)\exp(i\omega t)}^2\bigg|\theta}.
\end{align}
It is easy to show that
\begin{align}
S_-(\omega|\theta) &= S_A' + S_A + |\chi_-(\omega)|^2(S_B-S_A),
\label{S_minus}\\
S_+(\omega|\theta) &= S_A' + S_A + |\chi_+(\omega)|^2(S_B+S_A),
\label{S_plus}
\end{align}
where $\chi_\pm(\omega)$ are the transfer functions that depend on the
other known parameters. Since
\begin{align}
\frac{4|g|^2}{\gamma_a\gamma_b} \ll 1
\end{align}
in the experiment, $|\chi_+(\omega)|^2\approx |\chi_-(\omega)|^2$, and
the asymmetry of the two spectra can be attributed to the presence of
$S_A$, the input optical noise \cite{khalili2012,classical_model}.

Another statistic of interest is the steady-state mechanical energy:
\begin{align}
\lim_{t\to\infty} \expect\Bk{|b(t)|^2\big|\theta} \approx S_B.
\end{align}
With appropriate normalizations, the quantum theory will result in
the following constraints:
\begin{align}
S_A &\ge 0.5
\textrm{ and }
S_B \ge 0.5,
\end{align}
which are manifestations of the uncertainty principle for the optical
and mechanical quadratures.  Quantum gravity theories might violate or
modify the uncertainty principle, resulting in different constraints
\cite{pikovski,chen2013,blencowe}.  For example, a quantum gravity
theory may assume
\begin{align}
S_B \ge 0.5+\epsilon,
\label{modified}
\end{align}
where $\epsilon$ is a parameter that depends on the mechanical mass
\cite{pikovski}.

\subsubsection{\label{parallel}Parallel Kalman filters}
Statistics cannot be measured exactly in finite time, so let us turn
to the Bayesian approach to characterize the uncertainties.  Our first
task is to compute $P(Y|\theta)$ for many points that cover the
two-dimensional plane of $\theta = (S_A,S_B)^\top$. We can take
advantage of the Gauss-Markov property of the model and numerically
compute $P(Y|\theta)$ efficiently using the famous Kalman filter in a
multiple-model approach \cite{bar-shalom}. The procedure is as
follows:
\begin{enumerate}
\item For the first set of measurements and each $\theta$, define a
  normalized vectoral observation process as
\begin{align}
y_t &\doteq \sqrt{\frac{2}{S_A'+S_A}}\int_0^t d\tau 
\bk{\begin{array}{c} \real A_-(\tau)\\
 \imag A_-(\tau)
\end{array}},
\end{align}
such that the white noise in $y_t$ is normalized to give
\begin{align}
dy_t dy_{t}^{\top} = I dt,
\end{align}
with $I$ being the identity matrix.

\item Pass $y_t$ through a Kalman filter that assumes the same
  $\theta$ and Eqs.~(\ref{aminus})--(\ref{Aminus}). Specifically, let
\begin{align}
Y_t \doteq \BK{y_\tau, 0\le \tau\le t}
\end{align}
be the observation record up to time $t$, and 
\begin{align}
x_t \doteq \bk{\begin{array}{c}\real a(t)\\
\imag a(t)\\
\real b(t)\\
\imag b(t)
\end{array}} 
\end{align}
be the state vector. The Kalman filter
\cite{kalman,bar-shalom,simon,wiseman_milburn} is an algorithm that
determines the Gaussian posterior distribution $P(x_t|Y_t,\theta)$
given the past observation record $Y_t$ by computing its mean and
covariance matrix (see Appendices~\ref{dkalman} and \ref{ckalman} for
the formulas).

\item Combine the outputs from the Kalman filter with $y_t$ to obtain
  $P(Y_-|\theta)$ for the first set of measurements. In continuous
  time, the formula is \cite{hypothesis}
\begin{align}
&\quad P(Y_-|\theta) 
\nonumber\\
&= P_W(Y_-)
%\nonumber\\&\quad\times
\exp\Bk{\int_0^T dy_t^\top \mu_t(\theta)
-\frac{1}{2}\int_0^T dt \mu_t^\top(\theta) \mu_t(\theta)},
\end{align}
where $P_W(Y_-)$ is the probability measure of a vectoral Wiener
process (with zero increment mean and variance $dy_t dy_t^\top =
Idt$), $\mu_t(\theta)$ are the filtering estimates of the following
state variables:
\begin{align}
\mu_t(\theta) &\doteq
\sqrt{\frac{2}{S_A'+S_A}} \bk{
\begin{array}{c}\real \expect[\sqrt{\gamma_a}a(t)|Y_t,\theta]\\
\imag \expect[\sqrt{\gamma_a}a(t)|Y_t,\theta]
\end{array}},
\end{align}
which can be extracted from the Kalman filter estimates
$\expect(x_t|Y_t,\theta)$, and the $dy_t$ integral is an It\=o
integral, that is, $dy_t$ should be the increment ahead of $t$ and
$\mu_t(\theta)$ should not depend on $dy_t$.

Note that, in any computation of the posterior distribution of
$\theta$, $P_W(Y_-)$ appears in both the numerator and denominator of
the Bayes theorem and, being independent of $\theta$, cancels itself.

\item Repeat Step 1-3 for the second set of measurements to obtain
  $P(Y_+|\theta)$, assuming Eqs.~(\ref{aplus})--(\ref{Aplus}). $P(Y|\theta)$ is then
$P(Y_+|\theta)P(Y_-|\theta)$.

\item Repeat Step 1-4 for all possible $\theta$. 

\end{enumerate}
The tricky part is Step 5, as we need to set an appropriately large
but fine grid that discretizes $\theta$ in practice.  Fortunately, the
Kalman filters can be computed in parallel for different values of
$\theta$, so we can exploit parallel computing power to sweep many
$\theta$ values, until $P(Y|\theta)$ becomes relatively smooth inside
the considered region and negligible outside it.

\subsubsection{Expected information}
For a useful guide on how to construct the grid for the parallel
Kalman filters and also how well the signal processing technique is
expected to work, we can consult the information measures introduced
in Sec.~\ref{stat}. Consider, for example, two hypotheses with precise
assumed values for $\theta$. The problem then becomes a discrimination
between two vectoral, complex, stationary, zero-mean, and Gaussian
processes with power-spectral-density matrices
\begin{align}
S_0 &=
\bk{\begin{array}{cc}S_-(\omega|\theta_0)& 0\\
0 & S_+(\omega|\theta_0)\end{array}},
\\
S_1 &=
\bk{\begin{array}{cc}S_-(\omega|\theta_1)& 0\\
0 & S_+(\omega|\theta_1)\end{array}}.
\end{align}
The relative entropy and the Chernoff information have the following
long-time limits \cite{kazakos}:
\begin{align}
%&\quad 
\lim_{T\to\infty}\frac{D(P_1||P_0)}{T}
%\nonumber\\
&= \int \frac{d\omega}{2\pi}
\Bk{\trace S_0^{-1}\bk{S_1-S_0}-\ln|S_0^{-1}S_1|},
\\
%&\quad 
\lim_{T\to\infty}\frac{C(s)}{T}
%\nonumber\\
&= \int \frac{d\omega}{2\pi}
\ln \frac{|(1-s)S_0+s S_1|}{|S_0|^{1-s}|S_1|^s},
\label{C_asym}
\end{align}
where $|\cdot|$ is the determinant and the frequency integral should
be applied to only positive frequencies if the processes are
real. These expressions show that the information measures should
increase linearly with time asymptotically. The increase of
information with time is important, as it suggests that one can always
compensate for a bad signal-to-noise ratio by increasing the
measurement time.

For parameter estimation, the Chernoff information given by
Eq.~(\ref{C_asym}) can be used to compute the Cram\'er-Rao bound via
Eq.~(\ref{JC}) and the Ziv-Zakai bounds \cite{bell}. These
parameter-estimation bounds are especially useful for setting the grid
resolution for the parallel Kalman filters.

\subsubsection{Hypothesis testing and parameter estimation}
After the hard work of computing $P(Y|\theta) = P(Y|S_A,S_B)$, we can
now test the composite hypotheses about the uncertainty principle by
considering various $P(S_A,S_B|\mathcal H_j)$. First consider the
hypotheses used by Ref.~\cite{safavi2012}. One hypothesis assumes a
classical model with equal spectra for Eqs.~(\ref{S_minus}) and
(\ref{S_plus}), meaning that $S_A = 0$, and the other one assumes a
quantum model with $S_A = 0.5$. This implies
\begin{align}
P(S_A,S_B|\mathcal H_0) &= \delta(S_A)P(S_B|\mathcal H_0),
\\
P(S_A,S_B|\mathcal H_1) &= \delta(S_A-0.5)P(S_B|\mathcal H_1).
\end{align}
The difference in the assumed optical noise powers $S_A$ can make the
test favor one hypothesis over the other even if the data contain no
significant information about the mechanical mode (see, however,
Ref.~\cite{safavi_reply} for a different opinion).  It is obvious that
one can infer a lot more information about $S_B$ from the measurements
(as was done in Ref.~\cite{safavi2013a}), and the hypotheses should
make different assumptions about $S_B$, not $S_A$, if a test of the
mechanics is intended.

%One possibility is to partition the $(S_A,S_B)$
%plane into tiles with equal areas:
%\begin{align}
%&\quad P(S_A,S_B|\mathcal H_{jk}) 
%\nonumber\\
%&= 
%\Big\{\begin{array}{lll}
%1/\Delta^2, & j \le S_A/\Delta < j+1, & k \le S_B/\Delta < k+1,\\
%0, & \textrm{otherwise}.
%\end{array}
%\end{align}
%If we set $\Delta = 0.5$ for example, we can clearly identify the
%tiles that approach or violate the uncertainty principle and assign
%posterior probabilities to them. 

Without any obvious choice of $P(S_A,S_B|\mathcal H_j)$, we can also
treat the problem as parameter estimation using an objective prior
$P(S_A,S_B)$. The Jeffreys prior given by Eq.~(\ref{jeffreys}) can be
approximated by considering Eq.~(\ref{C_asym}) and using the identity
in Eq.~(\ref{JC}).  The posterior distribution is then
\begin{align}
P(S_A,S_B|Y) &= \frac{P(Y|S_A,S_B)P(S_A,S_B)}
{\int dS_A dS_B P(Y|S_A,S_B)P(S_A,S_B)},
\end{align}
which can be plotted graphically for visual impact, and a credible
region can be assigned according to Eq.~(\ref{credible}). To claim a
successful observation of zero-point mechanical motion, the whole
credible region should be close to $S_B = 0.5$.  When the mechanical
oscillator is very close to absolute zero, the credible region may
also be used to rule out modified uncertainty principles given by
Eq.~(\ref{modified}) by noting the values of $S_B$ that are well
outside the credible region.
%In addition to the Bayesian analysis, it
%is useful to report $P(Y|S_A,S_B)$, such that other scientists can
%draw their own conclusions based on their prior beliefs and other
%experimental evidence.

Using an atomic ensemble as the mechanical oscillator, Brahms
\textit{et al.}\ have performed an experiment \cite{brahms2012}
similar to the one we have studied. A careful analysis of this
experiment is left as an exercise for the reader.

\subsection{Caveat: systematic errors}
The Bayesian approach can perform worse than expected if the
model assumptions do not hold in practice. The errors due to wrong
assumptions are commonly called systematic errors. Here are a list of
possible sources:

\begin{enumerate}
\item \emph{Parameter uncertainties}. In our optomechanics example in
  Sec.~\ref{optomech}, we have assumed that some of the parameters,
  such as the resonance frequencies and the damping rates, are known
  exactly in advance. If not, one useful system identification method
  for prior calibration is the expectation-maximization (EM)
  algorithm, which is able to estimate most (not all) parameters of a
  homogeneous HMM \cite{shumway_stoffer}; see also Ref.~\cite{ang} for
  an application of the EM algorithm to an optomechanics
  experiment. If the parameters cannot be estimated exactly in
  advance, they would need to be included in $\theta$.

\item \emph{Parameter drifts}. A more serious problem occurs if the
  parameters are both unknown and drifting in time. Stationary
  statistics can no longer be assumed, and the Kalman filters cease to
  be optimal if the parameter drift is random. To deal with this, we
  have to take the parameters as part of the hidden state variables
  and perform nonlinear estimation. Optimal nonlinear estimation is
  extremely difficult to implement, but there exist many battle-tested
  approximations.  Methods based on Kalman filters include the
  extended and unscented Kalman filters \cite{simon}.  A notable
  example is the Gravity Probe B experiment, which relies on the
  unscented Kalman filter to perform the parameter estimation
  \cite{gravity_probe_b}.

\item \emph{Parameter ambiguities}. If there are too many unknown
  parameters, different combinations of the parameters may lead to the
  same $P(Y|\theta)$, and the data would not be able to distinguish
  such possibilities. Ignoring the alternatives may lead to serious
  actual errors and over-confidence in the estimates.

  To avoid committing this error, minimizing the number of unknown
  parameters helps tremendously. For simple models, this can be done
  by considering similarity transformations \cite{zhou,levy}, a
  technique for finding equivalent models that give the same
  observation statistics and discovering parameter
  redundancies.
  % Physicists may already be quite familiar with similarity
  % transformations without knowing it: they are called gauge
  % transformations in quantum field theory.

  The use of similarity transformations is especially important for
  the EM algorithm \cite{levy}, as the algorithm can be formulated to
  treat all parameters of a model as unknown and produce one set of
  estimates, ignoring all the other possibilities and giving one a
  false sense of certainty. If one is still left with too many
  parameters after careful considerations, independent calibrations
  and experiments to provide prior evidence for $P(\theta|\mathcal
  H_j)$ would be needed to narrow down the unknowns further.

\item \emph{Model mismatch}. Our model in Sec.~\ref{optomech} ignores
  the complication that the mechanical mode is coupled to another
  optical mode via laser cooling \cite{safavi2012}. This means that
  Eqs.~(\ref{bminus}) and (\ref{bplus}) are approximations. A
  higher-order HMM, that is, one with more state variables, is needed
  to model the actual situation more accurately, especially if there
  are other noticeable resonances in the data.

  A more troubling implication for fundamental physics tests is that
  the mechanical noise $B(t)$ actually has a significant optical
  origin due to the laser cooling.  If we already assume that an
  optical source must have $S_A \ge 0.5$, it would be inconsistent to
  assume that $S_B$ may go below $0.5$.  One needs to formulate the
  hypotheses much more carefully to avoid logical inconsistencies such
  as this.
\end{enumerate}
Systematic errors are ``unknown unknowns'': things we do not know we
don't know \cite{rumsfeld}. They are much harder to catch, and worse
still, ignoring them may result in misplaced confidence in one's
estimates. To deal with such errors, it is a good idea in general to
be conservative with the prior assumptions, use different inference
algorithms to cross-check the results, and perform independent
calibrations if possible.

% \subsection{Failure of imagination}
% For a poorly calibrated experiment, there may be too many unknown
% parameters in $\theta$. Different combinations of the parameters may
% then lead to the same observation probability $P(Y|\theta)$.
% This ambiguity may lead to the wrong conclusion if
% it is not properly accounted for.
% % This is the failure of imagination, one of the main reasons cited
% % for the failure in preventing the September 11 attack \cite{911}.

% 

%Failure of imagination is an honest mistake, but it is often used as
%an excuse for more cynical motives. One is the desire to generate
%hype.  By choosing a particular alternative model that deviates
%significantly from the expected hypothesis despite better
%alternatives, one can make the result look more surprising or
%convincing. Examples are too many to be cited here.

\subsection{Testing quantum jumps}
Let us come back to the mechanical oscillator and study its energy
dynamics. Under the linear model described in Sec.~\ref{optomech},
the equation of motion for the analytic signal in the absence of
measurements would be
\begin{align}
\diff{b(t)}{t} &= -\frac{\gamma}{2} b(t) + \sqrt{\gamma} B(t),
\end{align}
where we have suppressed the $b$ subscripts for clarity and will also
write $S = S_B$.  Consider the mechanical energy. Under classical
mechanics, it would be defined as
\begin{align}
\varepsilon(t) &= |b(t)|^2.
\end{align}
To derive an equation of motion for it, we should use stochastic
calculus. From It\={o} calculus, the result is
\begin{align}
d\varepsilon(t) &= -\gamma\Bk{\varepsilon(t) - S}dt
+\sqrt{2\gamma S\varepsilon(t)} dW(t),
\label{rayleigh}
\end{align}
where $dW(t)$ is a Wiener increment and models white noise
\cite{gardiner}.
%with the following
%properties:
%\begin{align}
%E\Bk{dW(t)|\mathcal E(t)} &= 0,
%\\
%E\Bk{dW^2(t)|\mathcal E(t)} &= dt,
%\end{align}
%and zero higher-order moments (relative to $dt$). $\mathcal E(t)$ is
%the history of $\varepsilon(\tau)$ previous to time $t$:
%\begin{align}
%\mathcal E(t) &\doteq \BK{\varepsilon(\tau),t_0\le \tau < t},
%\end{align}
%also called a filtration \cite{elliott}.

An alternative representation of the dynamics is the forward Kolmogorov
equation \cite{gardiner}:
\begin{align}
\parti{}{t} P(\varepsilon,t)
&= \int d\varepsilon' A(\varepsilon|\varepsilon') P(\varepsilon',t),
\label{kolmogorov}
\end{align}
also known as the Fokker-Planck equation or the master equation.  The
transition function $A(\varepsilon|\varepsilon')$, assuming the linear
model (designated as $\mathcal H_0$), is
\begin{align}
A(\varepsilon|\varepsilon',\mathcal H_0) 
&= 
\gamma\parti{}{\varepsilon}\delta(\varepsilon-\varepsilon')
\bk{\varepsilon'-S}
%\nonumber\\&\quad
+\gamma S\partit{}{\varepsilon}\delta(\varepsilon-\varepsilon')\varepsilon'.
\label{A0}
\end{align}
The steady-state distribution for $\varepsilon(t)$ is given by
\begin{align}
P_{\rm ss}(\varepsilon|\mathcal H_0) &= 
\frac{1}{S}\exp\bk{-\frac{\varepsilon}{S}},
\label{Pss0}
\end{align}
with moments
\begin{align}
\expect_{\rm ss}\Bk{\varepsilon^m|\mathcal H_0} &= m! S^m.
\end{align}
For example, the mean and variance are
\begin{align}
\bar\varepsilon_0 &\doteq \expect_{\rm ss}\Bk{\varepsilon|\mathcal H_0} = S,
\\
\overline{\Delta\varepsilon_0^2} &\doteq 
\expect_{\rm ss}\Bk{\bk{\varepsilon-\bar\varepsilon_0}^2\Big|\mathcal H_0} = S^2.
\label{var0}
%\\
%\overline{\Delta\varepsilon_0^3} &\doteq 
%\expect_{\rm ss}\Bk{\bk{\varepsilon-\bar\varepsilon_0}^3|\mathcal H_0}
%= 2S^3.
\end{align}
All the properties of the continuous energy model should be consistent
with the statistics of homodyne or heterodyne detection in an
optomechanics experiment; after all, all we have done is a change of
variables.

Eq.~(\ref{rayleigh}) predicts a continuous energy, whereas the quantum
theory can also result in a discrete energy model if we measure in the
phonon-number basis. Experimental progress towards such a measurement
in optomechanics is reported by Thompson \textit{et al.}\
\cite{thompson} and Sankey \textit{et al.}\ \cite{sankey}.  The
discrete jumps mean that it is difficult to write an equation of
motion that resembles Eq.~(\ref{rayleigh}), and it is more common to
represent the dynamics just by the forward Kolmogorov equation given
by Eq.~(\ref{kolmogorov}). For the damped quantum harmonic oscillator,
we have \cite{gardiner_zoller}
\begin{align}
A(\varepsilon|\varepsilon',\mathcal H_1) &= 
\delta(\varepsilon -1- \varepsilon')\Gamma_+(\varepsilon')
+\delta(\varepsilon +1 - \varepsilon')\Gamma_-(\varepsilon')
\nonumber\\&\quad
-\delta(\varepsilon-\varepsilon')\Bk{\Gamma_+(\varepsilon') 
+ \Gamma_-(\varepsilon')},
\label{A1}
\end{align}
where $\Gamma_\pm$ are the jumping rates:
\begin{align}
\Gamma_\pm(\varepsilon') &= \gamma(S\mp 0.5)(\varepsilon'\pm 0.5),
\end{align}
and $\varepsilon$ is restricted to discrete levels:
\begin{align}
\varepsilon \in \{0.5,1.5,\dots\}.
\end{align}
The steady state is
\begin{align}
P_{\rm ss}(\varepsilon|\mathcal H_1) &=
\frac{1}{S+0.5}\bk{\frac{S-0.5}{S+0.5}}^{\varepsilon-0.5},
\label{Pss1}
\end{align}
with the mean and variance given by
\begin{align}
\bar\varepsilon_1 &= S,
\\
\overline{\Delta\varepsilon_1^2} &= S^2-0.5^2.
\label{var1}
%\\
%\overline{\Delta\varepsilon_1^3} &= 2S(S^2-0.5^2).
\end{align}
In practice, $\varepsilon(t)$ is a hidden variable and observed
indirectly, so a measurement model should be constructed to include
observation noise and any measurement backaction effect.

Although Eq.~(\ref{A1}) is also a classical HMM, it is radically
different from the HGMM suitable for heterodyne and homodyne
detection, meaning that, in order to reproduce the quantum theory, the
classical models are \emph{contextual} with respect to the type of
measurement being performed.

Given the prior success of the linear model, the hypothesis given by
Eq.~(\ref{A0}) is a compelling alternative to Eq.~(\ref{A1}).
Evidence for the discrete energy model against the continuous
alternative would be a more direct confirmation of the original
quantal hypothesis and, together with the observation of the
uncertainty principle, a convincing demonstration of the quantum
probability theory for mechanics.

There are two ways of testing the discrete-energy hypothesis, both
difficult but in different ways. One is to sample $\varepsilon(t)$ at
very sparse time intervals, such that the samples can be assumed to be
independent and identically distributed (\textit{i.i.d.}), and the
test becomes a simple one between the steady-state distributions given
by Eqs.~(\ref{Pss0}) and (\ref{Pss1}). The statistical analysis is
relatively easy given the \textit{i.i.d.}\ property, but the procedure
is very inefficient especially if the observation noise is high, as it
throws away most of the data that can be obtained in-between the sampling
times.

A much more efficient method is to consider a continuous measurement
of $\varepsilon(t)$ and perform hypothesis testing on the whole record
and discriminate between Eqs.~(\ref{A0}) and (\ref{A1}) directly, as
proposed by Tsang~\cite{hypothesis}.  The statistical analysis becomes
much more complicated however, as the observation processes are highly
non-Gaussian.  Stochastic calculus helps \cite{hypothesis,mismatch},
but analytic results are more difficult to obtain than the ones for
the linear model in Sec.~\ref{optomech}. We leave this interesting
problem for future work.

Note that there are alternative approaches to testing quantum jumps
that are not based on statistical hypothesis testing
\cite{santamore,santamore_pra,jacobs2007,miao_prl,clerk_prl}.  A
critique of these methods is left as an exercise for the reader.

\subsection{Contextuality}
With the demonstrations of the uncertainty principle and the discrete
energy, the quantum proposition would become a lot more attractive: we
get two contextual classical models for the price of one.  Yet the
skeptics may still ask the following:
\begin{enumerate}
\item Is there a noncontextual classical model, beyond the
  representations we have considered, that can explain both phenomena,
  or all quantum phenomena in general?

\item Are two contextual models really that bad, if it means
  one can avoid the Hilbert-space theory?
\end{enumerate}

To address the first question, we can appeal to the
Bell-Kochen-Specker theorem, which is a no-go theorem that rules out
the possibility of one noncontextual classical model to explain all
quantum phenomena, if we impose certain restrictions on the classical
state variables \cite{peres,mermin}.  Of course, it is still a
fundamental open question whether an efficient classical description
of quantum mechanics is possible if we relax the restrictions
somewhat.

To address the second question, we can appeal to the power of quantum
computation: it is known that linear bosonic dynamics, together with
discrete-energy sources and measurements, is sufficient to perform
universal quantum computation \cite{klm} and solve difficult problems
\cite{aaronson} efficiently. This means that, if an experiment
performs operations that require \emph{switching} between the
different contexts, naive contextual models can fail, as it is not
even known if an efficient classical description exists at all.

\subsection{Nonlocality}
Contextuality is a serious inconvenience for classical models, but we
may also ask whether there are other more fundamental reasons for
finally giving up on classical mechanics. Bell's theorem and its
generalizations try to provide one by pitting classical mechanics
against special relativity: to reproduce the quantum theory, the
classical hidden variables must be able to communicate at superluminal
speeds \cite{peres}. Moreover, by providing explicit inequalities that
classical models with local hidden variables must obey, the Bell
theorems can be tested experimentally. The interested readers are
referred to more knowledgeable sources
\cite{peres,mermin,brunner,emary2013} on this topic; we emphasize only
that statistical hypothesis testing methods can and should be applied
to such tests, as proposed by Peres~\cite{peres_bayes} and van Dam
\textit{et al.}~\cite{vandam}.

Rather than focusing on the constraints and the no-go theorems, we
might ask a more positive question: does the contextuality or the
nonlocality of a quantum system confer any useful advantage in
information processing applications? It is perhaps this question that
inspired the emergence of quantum information science
\cite{nielsen,caves_qis}, and it is perhaps a critical examination of
this question that will ensure a sustainable development of the field.

\section{Conclusion}
As the take-home message, we conclude with the following quote:

\begin{quotation}
\emph{We've learned from experience that the truth will come
  out. Other experimenters will repeat your experiment and find out
  whether you were wrong or right. Nature's phenomena will agree or
  they'll disagree with your theory. And, although you may gain some
  temporary fame and excitement, you will not gain a good reputation
  as a scientist if you haven't tried to be very careful in this kind
  of work. And it's this type of integrity, this kind of care not to
  fool yourself, that is missing to a large extent in much of the
  research in Cargo Cult Science.}

\emph{... So I have just one wish for
  you---the good luck to be somewhere where you are free to maintain
  the kind of integrity I have described, and where you do not feel
  forced by a need to maintain your position in the organization, or
  financial support, or so on, to lose your integrity. May you have
  that freedom.}
---Richard P. Feynman \cite{cargo_cult}
\end{quotation}

\begin{acknowledgments}
  It is impossible to list all the people who help shape the views
  expressed here through various forms of interactions, but surely the
  most important are the ones who have also paid me while I indulge in
  these issues, including Carl Caves, Jeff Shapiro, Seth Lloyd,
  Demetri Psaltis, the various funding agencies, and the National
  University of Singapore. I also thank Kurt Jacobs, who provided some
  insightful suggestions, and my group members, who are the main
  motivation for my writing this tutorial.

%  The choice of topics presented here reflects my personal preferences
%  and incomplete knowledge and should not be taken as a comprehensive
%  review of all the subjects relevant to quantum mechanics
%  tests. Similarly, the choice of references is biased and not meant
%  to be a literature survey. As a result, if I did not discuss your
%  favorite topic or cite your favorite papers, I hope you understand.
%  For those who are keen to explore the literature further, the cited
%  textbooks and review papers should serve as good starting points.

  This work is supported by the Singapore National Research Foundation
  under NRF Grant No.~NRF-NRFF2011-07.
\end{acknowledgments}
\appendix

\section{\label{hmm}Hidden Markov models (HMM)}
An HMM expresses the probability function $P(Y)$ of an observed
variable $Y$ in terms of a hidden variable $X$ as follows:
\begin{align}
  P(Y) &= \sum_X P(Y,X),
\end{align}
with assumptions about how the variables are related to each other.
%Control theorists will be more familiar with the treatment in this
%section, which follows closely Ref.~\cite{elliott}, whereas physicists
%may be baffled by all the technical terms in Ref.~\cite{elliott}, so
%it is worth clarifying the difference here.

In the following I define the most basic type of HMM with discrete
time and discrete possibilities, following closely the treatment in
Ref.~\cite{elliott}.

\subsection{\label{state}State vector}
At each time, the system of interest is in \emph{one} of $N$ possible
states. A possibility is denoted by $n$, $n = 1,\dots,N$.  For
example, with $D$ bits, $N = 2^D$, and each $n$ denotes a particular
bit sequence. The state at time $k$ is represented by a vector
\begin{align}
x_k \in \BK{e^0,\dots,e^{N-1}}
\end{align}
in state space, and the global state vector $X$ is
\begin{align}
X = \BK{x_K,\dots,x_0}.
\end{align}
Here the superscripts are indices and should not be confused with
powers; the meaning should be clear given the context.  The unit
vectors $e^n$ represent the different possibilities of a state.  They
are in an $N$-dimensional Euclidean state space:
\begin{align}
e^0 = \bk{\begin{array}{c}1\\ 0\\ \vdots \\ 0\end{array}},\ 
e^1 = \bk{\begin{array}{c}0\\ 1\\ \vdots \\ 0\end{array}},\
\dots
\end{align}
and orthogonal to each other in terms of the inner product:
\begin{align}
\Avg{e^n, e^m} &= \delta^{nm} 
\doteq \Big\{\begin{array}{ll}
1, & n = m,\\
0, & \textrm{otherwise}.
\end{array}
\label{orthonormal}
\end{align}

\subsection{\label{functions}State functions}
It is important to emphasize that $x$ is an indicator of the
possibility and does not carry any other physical property.  A
property of a state can be quantified by a value $F^n$ assigned to
each possibility $n$. To write the value as a function $F(x)$, let
\begin{align}
\Avg{e^n, x} = 1_{e^n}(x) \in \{0,1\}
\end{align}
be the $n$th component of $x$, where $1_{e^n}(x)$ is an indicator
function:
\begin{align}
1_{\Xi}(x) &= \Big\{
\begin{array}{ll}
1, & x \in \Xi,
\\
0, & \textrm{otherwise}.
\end{array}
\label{indicator}
\end{align}
$\Avg{e^n,x}$ is a binary (``yes-no'') variable that indicates whether
$x$ is in state $n$. We can then write
\begin{align}
  F(x) &= \sum_n F^n \Avg{e^n,x}.
\label{function}
\end{align}
Note the subtle difference between a function $F(x)$ and its possible
values $F^n$.  For example, the identity function is
\begin{align}
Ix &= \sum_n  e^n \Avg{e^n,x} = x,
\end{align}
and when multiplied by a matrix $A$,
\begin{align}
Ax &= \sum_{n,m}   A^{nm} e^n\Avg{e^m,x}.
\end{align}
We see that the definition of a state function here depends heavily on
the assumption that the system is always in one and only one of the
possible states.

\subsection{Initial probability function}
At time $k = 0$, a nonnegative probability $P_0^n$ is assigned to each
$e^n$. The probability \emph{function} of a state variable $x_0$ is
written as
\begin{align}
P(x_0) &= \sum_n   P_0^n \Avg{e^n,x_0}.
\label{initial_P}
\end{align}
The probability distribution can be extracted from the function by
\begin{align}
P_0^n  &= P(x_0=e^n).
\end{align}
Note the subtle difference between the function $P(x_0)$ and the
distribution $P_0^n$.

\subsection{Markovianity}
The state described by $X$ is hidden and inferred only through an
observed variable $Y$. Similar to $X$, $Y$ can also be broken down
into a series of observation state vectors
\begin{align}
y_k &\in \BK{f^1,\dots,f^M},
\\
Y &= \BK{y_K,\dots,y_1},
\end{align}
where $f^n$ are unit vectors similar to $e^n$.  
Define
\begin{align}
Y_k &\doteq \BK{y_k,\dots,y_1},
\\
X_k &\doteq \BK{x_k,\dots,x_0}.
\end{align}
The Markov property assumes that $x_{k+1}$ and $y_{k+1}$ depend only
on the previous $x_{k}$, such that
\begin{align}
P(y_{k+1},x_{k+1}|Y_k,X_k) &= P(y_{k+1},x_{k+1}|x_{k}),
\end{align}
which leads to
\begin{align}
P(Y,X) &= P(y_K,x_K|x_{K-1})\dots P(y_1,x_1|x_0)P(x_0).
\end{align}
It is common to assume that the system noise and the observation 
noise are independent:
\begin{align}
P(y_{k+1},x_{k+1}|x_{k}) &= P(y_{k+1}|x_{k}) P(x_{k+1}|x_{k}),
\end{align}
although this is often not the case in classical models of quantum
optics.

We now have the complete specification of an HMM, and in principle we
can use it to calculate any multi-time statistic by taking the
expectation of any function $F(X,Y)$. For further details, more
general HMM, and their applications, see Ref.~\cite{elliott}.

%\section{\label{cest}Bayesian estimation}
\subsection{\label{cfilter}Bayesian filtering}
For simplicity, in the following we use the same notation to denote
probability functions and probability distributions. For example,
$P(x_k=e^n)$ is written as $P(x_k)$, and $\sum_{x_k}P(x_k)$ is taken
to mean $\sum_n P(x_k=e^n)$.

Bayesian filtering is a signal-processing technique that computes the
posterior distribution $P(x_k|Y_k)$ given the immediate past record
$Y_k$.  For an HMM, we can obtain a recursive formula via the
following:
\begin{align}
P(x_{k+1}|Y_{k+1}) &= \frac{P(x_{k+1},Y_{k+1})}{P(Y_{k+1})}
\\
&=\frac{\sum_{x_k} P(x_{k+1},x_k,y_{k+1},Y_{k})}{P(Y_{k+1})}
\\
&= \frac{\sum_{x_k} P(x_{k+1},x_k,y_{k+1}|Y_{k})P(Y_k)}{P(Y_{k+1})}
\\
&= \frac{\sum_{x_k}P(y_{k+1},x_{k+1}|x_k,Y_k)P(x_k|Y_k)}{P(y_{k+1}|Y_k)}
\\
&= \frac{\sum_{x_k}P(y_{k+1},x_{k+1}|x_k)P(x_k|Y_k)}{P(y_{k+1}|Y_k)}.
\label{filter_recursion}
\end{align}
In other words, $P(x_{k+1}|Y_{k+1})$ is obtained by starting from the
initial condition $P(x_0)$, propagating $P(x_k|Y_k)$ forward using
$P(y_{k+1},x_{k+1}|x_k)$, and normalizing the resulting
expression. Continuous-time limits of the filtering equation can be
found in Refs.~\cite{liptser,liptser2,smooth_pra1,smooth_pra2}.

\subsection{\label{csmoother}Bayesian smoothing}
The goal of Bayesian smoothing is to compute the posterior
distribution $P(x_k|Y)$ of the hidden state at a certain time in the
past given the complete observation record $Y$. It is usually more
accurate than filtering when $x_k$ is a stochastic process, as the
future record can contain information about $x_k$ that is not
available in the past, but it is less useful for real-time
applications that require information about the current and the
future, such as aircraft control and financial trading.

One method of smoothing
is to split $Y$ into the past record $Y_k = \{y_k,\dots,y_1\}$ and the
future record
\begin{align}
\bar Y_k = Y \setminus Y_k = \BK{y_K,\dots,y_{k+1}}.
\end{align}
We then have
\begin{align}
P(x_k|Y) &= P(x_k|\bar Y_k,Y_k)
\\
&= \frac{P(x_k,\bar Y_k|Y_k)}{P(\bar Y_k|Y_k)}
\\
&= \frac{P(\bar Y_k|x_k,Y_k)P(x_k|Y_k)}{P(\bar Y_k|Y_k)}
\\
&= \frac{P(\bar Y_k|x_k)P(x_k|Y_k)}{P(\bar Y_k|Y_k)}.
\label{symmetric}
\end{align}
In other words, $P(x_k|Y)$ is equal to the product $P(\bar
Y_k|x_k)P(x_k|Y_k)$ with normalization.  $P(x_k|Y_k)$ can be computed
using filtering, while $P(\bar Y_k|x_k)$ is given by
\begin{align}
&\quad P(\bar Y_k|x_k) 
\nonumber\\
&= \sum_{x_{k+1}}P(\bar Y_{k+1},y_{k+1},x_{k+1}|x_k)
\\
&= \sum_{x_{k+1}}P(\bar Y_{k+1}|y_{k+1},x_{k+1},x_k)P(y_{k+1},x_{k+1}|x_k)
\\
&= \sum_{x_{k+1}}P(\bar Y_{k+1}|x_{k+1})P(y_{k+1},x_{k+1}|x_k),
\label{backward}
\end{align}
which defines a backward-time recursion analogous to
Eq.~(\ref{filter_recursion}), starting from the final-time condition
$P(\bar Y_{K}) = 1$. Continuous-time limits of the smoothing equations
can be found in Refs.~\cite{pardoux,smooth_pra1,smooth_pra2}.

\subsection{\label{curse}Curse of dimensionality}
The principal difficulty with implementing Bayesian filtering and
smoothing in practice is that a probability distribution of $x_k$ is
specified by $O(N)$ numbers, and $N$ grows exponentially with the
degree of freedom $D$ in a system. This makes any direct computation
of an $N$-dimensional probability distribution extremely expensive for
large $D$; a problem known as the curse of dimensionality. One central
goal of statistical inference research is to find algorithms that
approximate a probability distribution using far less numbers
(relative to $D$), finish in a reasonable time (relative to the number
of time steps $K$ in the model), and still achieve acceptable
estimation performances.

\section{\label{hgmm}Hidden Gauss-Markov models (HGMM)}
\subsection{Discrete-time HGMM}
Suppose now that $x_k$ and $y_k$ are vectors of unbounded continuous
random variables:
\begin{align}
x_k &= \bk{\begin{array}{c}x_k^{(0)}\\ x_k^{(1)}\\ \vdots\end{array}}
\in \mathbb R^D,
&
y_k &= \bk{\begin{array}{c}y_k^{(0)}\\ y_k^{(1)}\\ \vdots\end{array}}
\in \mathbb R^{d}.
\end{align}
If the initial $P(x_0)$ and the transitional $P(y_{k+1},x_{k+1}|x_k)$
are Gaussian:
\begin{align}
&P(x_0) \propto \frac{1}{\sqrt{\det\Sigma_0}}
\exp\Bk{-\frac{1}{2}\bk{x_0-x_0'}^\top\Sigma_0^{-1}\bk{x_0-x_0'}},
\\
&\quad P(y_{k+1},x_{k+1}|x_k)
\nonumber\\
&\propto
\frac{1}{\sqrt{\det\Lambda_k}}
%\nonumber\\&\quad\times
\exp\Bk{-\frac{1}{2}\bk{z_{k+1}-\bar z_{k}}^\top\Lambda_k^{-1}
\bk{z_{k+1}-\bar z_{k}}},
\end{align}
with
\begin{align}
z_{k+1} &= \bk{\begin{array}{c}x_{k+1}\\ y_{k+1}\end{array}},
\quad
\bar z_{k} = 
\bk{\begin{array}{c}A_k x_{k} + B_k u_k\\ C_k x_{k}\end{array}},
\\
\Lambda_k &= 
\bk{\begin{array}{cc}Q_k & S_k\\ S_k^\top & R_k\end{array}},
\end{align}
the model is known as a hidden Gauss-Markov model (HGMM), which has
been extensively studied due to its analytic and computational
tractability. A more common representation is to define zero-mean
Gaussian system and observation noises as
\begin{align}
w_k &\doteq x_{k+1} - A_k x_k - B_k u_k,
\\
v_k &\doteq y_{k+1}- C_k x_k,
\end{align}
such that the equations of motion can be written as
\begin{align}
x_{k+1} &= A_k x_k + B_k u_k + w_k,
\label{hgmm_x}
\\
y_{k+1} &= C_k x_k + v_k,
\label{hgmm_y}
\end{align}
with noise statistics given by
\begin{align}
\expect\bk{w_k} &= \expect\bk{v_k} = 0,
\\
\expect\bk{w_kw_k^\top} &=Q_k,
\\
\expect\bk{v_k v_k^\top} &= R_k,
\\
\expect\bk{w_k v_k^\top} &=S_k.
\end{align}

\subsection{\label{dkalman}Kalman filter}
The Kalman filter \cite{kalman,bar-shalom,simon} is an algorithm that
computes the mean
\begin{align}
x_k' &\doteq \expect\bk{x_{k}|Y_{k}}
\end{align}
 and covariance matrix
\begin{align}
\Sigma_k &\doteq \expect\Bk{\bk{x_{k}-x_k'}\bk{x_{k}-x_k'}^\top|Y_{k}}
\end{align}
of the Gaussian posterior distribution given the immediate past observation
record $Y_k$ for the HGMM.  One trick of deriving the filter for
nonzero $S_k$ is to rewrite Eqs.~(\ref{hgmm_x}) and (\ref{hgmm_y}) as
\cite{bar-shalom}
\begin{align}
x_{k+1} &= A_k x_k + B_k u_k + w_k + T_k\bk{y_{k+1}-C_kx_k-v_k}
\\
&= \bk{A_k-T_kC_k}x_k + B_ku_k + T_k y_{k+1} + \xi_k,
\label{newx}
\\
y_{k+1} &= C_k x_k + v_k,
\label{newy}
\end{align}
where the redefined system noise is
\begin{align}
\xi_k &\doteq w_k - T_k v_k,
\end{align}
which can be made independent of $v_k$ if we set
\begin{align}
T_k &= S_kR_k^{-1},
\\
\expect\bk{\xi_kv_k^\top} &= 0,
\\
\expect\bk{\xi_k\xi_k^\top} &= Q_k - S_kR_k^{-1} S_k^\top.
\end{align}
This allows us to apply the Kalman filter for uncorrelated noises
to Eqs.~(\ref{newx}) and (\ref{newy}).
The result is
\begin{align}
\Gamma_k &= \Sigma_k C_k^\top\bk{C_k\Sigma_k C_k^\top + R_k}^{-1},
\label{kalman1}
\\
x_k'^+ &\doteq \expect\bk{x_k|Y_{k+1}} 
=  x_k' + \Gamma_k(y_{k+1}-C_kx_k'),
\label{kalman2}
\\
\Sigma_k^+ &\doteq \expect\Bk{\bk{x_k-x_k'^+}\bk{x_k-x_k'^+}^\top|Y_{k+1}}
\nonumber\\
&= \bk{I- \Gamma_k C_k}\Sigma_k,
\label{kalman3}
\\
x_{k+1}' &= A_k x_k'^+ + B_k u_k + S_kR_k^{-1}(y_{k+1}-C_kx_k'^+),
\label{kalman4}
\\
\Sigma_{k+1} &= (A_k-S_kR_k^{-1}C_k) \Sigma_k^+ (A_k-S_kR_k^{-1}C_k)^\top 
\nonumber\\&\quad
+ Q_k-S_kR_k^{-1}S_k^\top.
\label{kalman5}
\end{align}
The exceptional computational efficiency of the Kalman filter has made
it the standard filtering algorithm in engineering; many practical
filtering algorithms for non-Gaussian models, such as the extended and
unscented Kalman filters \cite{simon}, are based on HGMM
approximations and extensions of the Kalman filter.

\subsection{\label{dsmoother}Rauch-Tung-Striebel (RTS) smoother}
An HGMM smoother computes the mean and covariance of the Gaussian
posterior distribution given the whole observation record $Y$:
\begin{align}
\check x_k &\doteq \expect\bk{x_k|Y},
\\
\Pi_k &\doteq \expect\Bk{\bk{x_k-\check x_k}\bk{x_k-\check x_k}^\top|Y}.
\end{align}
The Rauch-Tung-Striebel (RTS) smoother \cite{rts,bar-shalom,simon} is
the most convenient algorithm. It first runs the Kalman filter given
by Eqs.~(\ref{kalman1})--(\ref{kalman5}) to obtain the set
$\{x_k'^+,\Sigma_k^+,x_{k+1}',\Sigma_{k+1},k=0,\dots,K-1\}$.  Then,
starting from
\begin{align}
\check x_K &= x_{K}',
&
\Pi_K &= \Sigma_K,
\\
\check x_{K-1} &= x_{K-1}'^+,
&
\Pi_{K-1} &= \Sigma_{K-1}^+,
\end{align}
the following formulas are iterated backward in time:
\begin{align}
\Upsilon_k &= \Sigma_k^+(A_k-S_kR_k^{-1}C_k)^\top\Sigma_{k+1}^{-1},
\\
\check x_k &= x_k'^+ + \Upsilon_k\bk{\check x_{k+1}-x_{k+1}'},
\\
\Pi_k &= \Sigma_k^+-\Upsilon_k(\Sigma_{k+1}-\Pi_{k+1})\Upsilon_k^\top.
\end{align}
For other forms of HGMM filters and smoothers, see
Refs.~\cite{bar-shalom,simon}.

\subsection{Continuous-time HGMM}
Define time as
\begin{align}
t_k &\doteq t_0 + k\delta t,
\end{align}
where $\delta t$ is the time interval between 
consecutive time steps. Suppose
\begin{align}
A_k-I &= f_k \delta t + o(\delta t),
\\
B_k &= b_k\delta t + o(\delta t),
\\
C_k &= c_k\delta t + o(\delta t),
\\
Q_k &= q_k\delta t + o(\delta t),
\\
R_k &= r_k \delta t + o(\delta t),
\\
S_k &= s_k\delta t + o(\delta t),
\end{align}
where $o(\delta t)$ denotes terms asymptotically smaller than $\delta t$.
We can then define the continuous-time limit of an HGMM in terms of
the following stochastic equations of motion:
\begin{align}
dx_t &= f_t x_t dt + b_t u_t dt +dw_t,
\\
dy_t &= c_t x_t dt + dv_t,
\end{align}
with noise properties given by
\begin{align}
\expect\bk{dw_t} &= \expect\bk{dv_t} = 0,
\\
\expect\bk{dw_tdw_t^\top} &=q_t dt,
\\
\expect\bk{dv_tdv_t^\top} &= r_tdt,
\\
\expect\bk{dw_t dv_t^\top} &=s_t dt.
\end{align}

\subsection{\label{ckalman}Kalman-Bucy filter}
The continuous-time limit of the Kalman filter in
Appendix~\ref{dkalman} is known as the Kalman-Bucy filter
\cite{kalman_bucy}. It is given by \cite{kalman_bucy,bar-shalom,simon}
\begin{align}
\Gamma_t &= \bk{\Sigma_t c_t^\top + s_t}r_t^{-1},
\label{kb1}
\\
dx_t' &= \bk{f_t x_t' + b_tu_t}dt + \Gamma_t(dy_t - c_t x_t' dt),
\label{kb2}
\\
\diff{\Sigma_t}{t} &= 
f_t\Sigma_t + \Sigma_t f_t^\top + q_t- \Gamma_t r_t\Gamma_t^\top.
\label{kb3}
\end{align}
This limit is useful for deriving analytic results and simplifying the
filter implementation, as the differential equations are easier to solve
analytically. 

\subsection{\label{mfp}Mayne-Fraser-Potter smoother}
Although there exists a continuous-time version of the RTS smoother
\cite{rts}, a time-symmetric form of the optimal smoother due to Mayne
\cite{mayne} and Fraser and Potter \cite{fraser} is more amenable to
analytic calculations.  It involves running the following filter,
which has the same form as the Kalman-Bucy filter, backward in time:
\begin{align}
\Gamma_t &= \bk{\Sigma_t c_t^\top + s_t}r_t^{-1},
\\
-dx_t'' &= -\bk{f_t x_t'' +b_t u_t} dt +\Gamma_t(dy_t-c_t x_t'' dt),
\label{xb}\\
-\diff{\Phi_t}{t} &= 
-f_t\Phi_t -\Phi_t f_t^\top + q_t- \Gamma_t r_t\Gamma_t^\top,
\label{Xi}
\end{align}
and combining the results with the forward Kalman-Bucy filter given by
Eqs.~(\ref{kb1})--(\ref{kb3}) as follows:
\begin{align}
\Pi_t &= \bk{\Sigma_t^{-1} + \Phi_t^{-1}}^{-1},
\\
\check x_t &= \Pi_t\bk{\Sigma_t^{-1}x_t' + \Phi_t^{-1}x_t''}.
\end{align}
The final-time conditions for Eqs.~(\ref{xb}) and (\ref{Xi}) should
correspond to $\Phi_T^{-1} = 0$. In practice, one can solve for
$\Phi_t^{-1}$ and $\Phi_t^{-1}x_t''$ instead of $\Phi_t$ and $x_t''$ to
avoid the ill-defined final-time conditions \cite{mayne,fraser}.

%We now skip all the boring details of measure changes and jump
%directly into a hidden Gauss-Markov model (HGMM):

%complex trick: turn all $^\top$ into $^\dagger$.

%\subsection{\label{ratio}Likelihood ratio}

%\subsection{Smoothers}

%\subsection{\label{em}Expectation-maximization (EM) algorithm}

%\section{Non-Gaussian hypothesis testing}
%\subsection{Likelihood ratio}

%\subsection{Mutual information and mean-square errors}

%\subsection{Relative entropy and mismatched mean-square errors}

\section{\label{qpt}Quantum probability theory}
\subsection{Hilbert space}
Consider an $N$-dimensional Hilbert space spanned by an
orthonormal basis
\begin{align}
\mathcal B_\phi &= \BK{\phi^0,\dots,\phi^{N-1}}.
\label{Phi}
\end{align}
For example, $N = 2^D$ for $D$ qubits. $\phi^n$ is a projection
operator, and in the bra-ket notation, it would be written as
\begin{align}
\phi^n &\equiv \ket{\phi^n}\bra{\phi^n},
\end{align}
where the $\equiv$ sign here means different notations for the same
quantity. The Hilbert-Schmidt inner product is written as
\begin{align}
\Avg{\phi^n,\phi^m} \equiv 
\trace\Bk{\bk{\ket{\phi^{n}}\bra{\phi^n}}^\dagger \ket{\phi^m}\bra{\phi^m}}
 = \delta^{nm}.
\end{align}
In classical probability theory, we assume that a state vector can
only be one of the unit vectors in one basis in a Euclidean space. The
key to quantum probability theory is that any basis in the Hilbert
space can be used to specify the possibilities.

\subsection{\label{qstate}Quantum state}
Consider a basis $\mathcal B_\xi$. Similar to the classical case, we can
define a state as one of \emph{its} possibilities:
\begin{align}
\psi \in \mathcal B_\xi,
\end{align}
such that the indicator function of $\psi$ with respect to
$\mathcal B_\xi$ is
\begin{align}
1_{\mathcal B_\xi}(\psi) &= 1.
\end{align}
$\psi$ is called a quantum state. In the bra-ket notation, we may
write it as
\begin{align}
\psi &\equiv \ket{\psi}\bra{\psi}.
\end{align}
Here $\psi\in \mathcal B_\xi$ implies that $\psi$ is compatible with
the basis $\mathcal B_\xi$, meaning that the state becomes equivalent
to a classical state if we restrict ourselves to state operations
within this basis.  Conversely, given any $\psi$, one can always find
a compatible basis $\mathcal B_\xi$ in which $\psi$ is an element.
This is a subtle but important point: it allows us to associate any
quantum state $\psi$ with a classical state of reality in the
\emph{context} of a compatible basis.

The $n$th component of $\psi$ in a compatible basis is
\begin{align}
\Avg{\xi^n,\psi} = 1_{\xi^n}(\psi) \in \{0,1\},
\label{qindicator}
\end{align}
which is a qualified indicator function like Eq.~(\ref{indicator}),
but for any basis in general, the inner product
\begin{align}
\Avg{\phi^n,\psi} \equiv \abs{\Avg{\phi^n|\psi}}^2
\end{align}
has the following properties
\begin{align}
0 \le \Avg{\phi^n,\psi} &\le 1,
\\
\sum_{n} \Avg{\phi^n,\psi} &= 1,
\end{align}
which hint at the role of $\Avg{\phi^n,\psi}$ as a probability function.

\subsection{Unitary maps}
An important class of operations on a quantum state are the unitary
maps, written as
\begin{align}
\mathcal U\psi \equiv U\ket\psi\bra\psi U^\dagger,
\end{align}
which models the transition from one state to another. The unitary
operator $U$ is expressed as
\begin{align}
U &= \sum_{n,m} U^{nm} \ket{\xi^n}\bra{\xi^m} = \sum_m\ket{\phi^m}\bra{\xi^m},
\\
\ket{\phi^m} &= \sum_n U^{nm}\ket{\xi^n} = U\ket{\xi^m},
\end{align}
where $U^{nm}$ is the unitary matrix that defines $U$:
\begin{align}
U^{nm} &= \bra{\xi^n}U\ket{\xi^m} = \Avg{\xi^n|\phi^m}.
\end{align}
Note the subtle difference between an operator and a matrix.

A special class of unitary operators is the permutation, which simply
assigns one ket to another in the same basis:
\begin{align}
U^{nm} &= \delta^{n,\pi(m)},
\\
\Avg{\xi^n,\phi^m} &= |U^{nm}|^2 \in\{0,1\}.
\label{permutation}
\end{align}
If $\psi \in \mathcal B_\xi$, a permutation would stay in the same basis, and
the transition becomes equivalent to a classical state transition.

In the other extreme, the Fourier-transform unitary assigns a state in
one basis to another in a ``maximally incompatible'' basis:
\begin{align}
U^{nm} &= \frac{1}{\sqrt{N}}\exp \frac{i2\pi nm}{N},
\\
\Avg{\xi^n,\phi^m} &= |U^{nm}|^2 = \frac{1}{N},
\label{mub}
\end{align}
which is useful for quantum computation \cite{nielsen}. Two bases that
satisfy Eq.~(\ref{mub}) are also called mutually unbiased
\cite{wootters_mub}.

To model continuous-time evolution, the unitary operator is expressed
in terms of a Hamiltonian operator $H$ as
\begin{align}
U(t) &= \exp(-iHt).
\end{align}

\subsection{von Neumann measurement}
Similar to the classical case, we can define a conditional probability
function with respect to two von Neumann measurements. A von Neumann
measurement is defined with respect to a basis $\mathcal B_\phi$, with
each outcome corresponding to a $\phi^n$. For one measurement in basis
$\mathcal B_\phi$ followed by another in $\mathcal B_\xi$, the
probability function of an outcome $\psi \in \mathcal B_\xi$,
conditioned on the previous outcome $\psi_0 \in \mathcal B_\phi$, is
\begin{align}
P(\psi|\psi_0) = \Avg{\psi,\psi_0} \equiv \abs{\Avg{\psi|\psi_0}}^2,
\label{born}
\end{align}
which is Born's rule. We can also model time evolution before the
final measurement by including a unitary map:
\begin{align}
P(\psi|\psi_0) = \Avg{\psi,\mathcal U\psi_0}.
\label{qm_nutshell}
\end{align}
Eq.~(\ref{qm_nutshell}) is quantum mechanics in a nutshell.
% Note how we completely avoided a discussion of observables (``there
% is no spoon'' \cite{matrix}).

A trivial but powerful property of Eq.~(\ref{qm_nutshell}) is unitary
invariance:
\begin{align}
P(\psi|\psi_0) &= \Avg{\mathcal U_0^* \psi,\mathcal U_0^*\mathcal U\psi_0},
\end{align}
where $\mathcal U_0$ is any unitary map. For example, if we let
$\mathcal U_0 = \mathcal U$, and since the adjoint is the same as the
inverse for a unitary, we obtain
\begin{align}
P(\psi|\psi_0) &= \Avg{\mathcal U^* \psi,\psi_0},
\end{align}
which is the Heisenberg picture. Any new picture can be generated by
choosing a $\mathcal U_0$, akin to a change of reference frame in
relativity. The interaction picture is a useful example.

In principle, Eq.~(\ref{qm_nutshell}) is all we need to compute
quantum probabilities, but it is extremely difficult to do so in
practice without further approximations if the degree of freedom $D$
is large. In the following, we consider the theoretical tools that can
facilitate this task.

\section{\label{quasi}Quasiprobability functions}
\subsection{Quantum-mechanics-free model}
If we restrict state operations (including the initial state, state
transitions, and measurements) to unit vectors in one basis, then the
quantum model becomes equivalent to a classical model without any
quantum feature, such as the uncertainty relations or measurement
invasiveness (this is called a quantum-mechanics-free model in
Ref.~\cite{qmfs}; see also Refs.~\cite{koopman,peres,gough}). It is,
however, possible to relax this restriction significantly and still
find a classical representation, if we incorporate probabilities. The
next sections describe how this can be done via Wigner functions.
%One naive way is to take $R$ incompatible bases and
%treat them as $R$ objects, resulting in an $N^R$-dimensional classical
%state space. This representation is very inefficient, however, and we
%can do much better if we incorporate probabilities.

\subsection{Mutually unbiased bases}
To pick the Hilbert-space bases for classical modeling, we start with
one, say,
\begin{align}
\mathcal B_{q} &= \BK{q^0,\dots,q^{N-1}},
\end{align}
and try to find all the bases that are unbiased with $\mathcal B_q$ and each
other according to Eq.~(\ref{mub}). We choose mutually unbiased bases
mainly because of the mathematical symmetry; there are some
practical benefits but we will not dwell on them for now.

If $N$ is a prime power, there are $R = N+1$ such bases including
$\mathcal B_q$ \cite{wootters_mub}. Let us focus on a prime $N$, and denote the
mutually unbiased bases by
\begin{align}
\tilde{\mathcal B} &= 
\BK{\mathcal B_{0},\mathcal B_1,\dots,\mathcal B_N},
\quad \mathcal B_N = \mathcal B_q,
\\
\mathcal B_r &= \BK{p_r^0,\dots,p_r^{N-1}},
\quad
r = 0,\dots,N-1.
\end{align}
For $N = 2$, the bases simply consist of the eigenstates of the three
Pauli operators. For the other primes, $\mathcal B_r$ can be
constructed from $\mathcal B_q$ using the fractional Fourier
transform.  We assume that one is interested only in state operations
with $\tilde{\mathcal B}$.

%\begin{align}
%\ket{p_r^n} &= \frac{1}{\sqrt{N}}\sum_m 
%\exp \frac{i2\pi(nm+rm^2)}{N}\ket{q^m}.
%\end{align}

The next step is to map the composite basis $\tilde{\mathcal B}$ to a
classical state space. A naive way would be to consider each $\mathcal
B_r$ to be a separate object; for example, a qubit ($N=2$) would be
modeled as three classical bits that correspond to the three spin
components.  This is obviously not the most efficient representation,
as the statistics of the classical bits must be correlated to model
one qubit.  In general, this naive approach would require an extremely
large $N^{N+1}$-dimensional classical state space.  Surprisingly, it
turns out that an $N^2$-dimensional classical state space is
sufficient, if we define an appropriate quasiprobability function in
analogy with the Wigner function for continuous variables
\cite{wootters_ap}.

\subsection{\label{dwigner}Discrete Wigner function}
Let $z$ be a classical state in one of $N^2$ possibilities. The
possibilities can be assigned to $N\times N$ points on a
two-dimensional lattice known as the phase space. For each $q^n$, we
assign a vertical line of classical states, denoted by the set
$\lambda(q^n)$. For the Fourier-transform basis $\mathcal B_0$, the
function $\lambda(p_0^n)$ also assigns a horizontal line of classical
states for each $p_0^n$.  Beyond the vertical and horizontal lines,
the basic idea of Ref.~\cite{wootters_ap} is to define tilted lines on
a lattice appropriately and construct a function $\lambda(p_r^n)$ that
provides a general mapping from any $p_r^n \in \tilde{\mathcal B}$ to
a line of classical states in the phase space. The discrete Wigner
function, defined via an operator $w(z)$ as
\begin{align}
W_0(z) &\doteq \Avg{w(z),\psi_0},
\end{align}
is then required to give the correct probability function that
coincides with Born's rule in Eq.~(\ref{born}) for any measurement in
any $\mathcal B_r \in \tilde{\mathcal B}$:
\begin{align}
  P(\psi|\psi_0) &= \sum_{z} 1_{\lambda(\psi)}(z) W_0(z)
\textrm{ for  } \psi \in \tilde{\mathcal B}.
\end{align}
A $w(z)$ that satisfies these properties for prime $N$ is reported in
Ref.~\cite{wootters_ap}.  If $N$ is not a prime, it can be factored
into a product of primes, and the procedure can be applied to a tensor
product of smaller Hilbert spaces with the prime dimensions.

Any quantum state transition within $\tilde{\mathcal B}$ can be represented by
an appropriate conditional probability function $W(z|z_0)$ in the
classical state space, such that Eq.~(\ref{qm_nutshell}) becomes
\begin{align}
P(\psi|\psi_0) &= \sum_{z,z_0}1_{\lambda(\psi)}(z) W(z|z_0)W_0(z_0)
\textrm{ for }
\psi \in \tilde{\mathcal B}.
\end{align}
As long as $\psi_0$ is also in $\tilde{\mathcal B}$, $W_0(z)$ is nonnegative,
and the quantum system can be modeled by a classical HMM with $N^2$
possible states.  However, if $\psi_0$ is not in $\tilde{\mathcal B}$ or if
$\mathcal U$ induces state transitions beyond the composite basis,
then $W_0(z)$ or $W(z|z_0)$ may become negative somewhere, hence the
name quasiprobability functions.

If $N$ is a prime power, the $N+1$ mutually unbiased bases can be used
to form the composite basis $\tilde{\mathcal B}$ directly, and
alternative Wigner functions can be defined with respect to
measurements in such bases without going through the composition; see
Ref.~\cite{gibbons}. For a discussion of the relationships between
nonnegative quasiprobability functions, contextuality, and quantum
information science in general, see
Refs.~\cite{ferrie_rpp,veitch2012,veitch2013}.

\subsection{\label{wigner}Wigner function for continuous variables}
The Wigner function was, of course, originally invented for unbounded
continuous variables, such as the position and momentum of a
mechanical oscillator and the quadratures of an optical field.  Its
properties and applications have been exhaustively studied; see, for
example, Refs.~\cite{hillery,walls_milburn,bartlett2012}.

The symmetry properties of the Wigner function is extremely powerful
for modeling a large class of quantum operations with minimal
contextuality.  In particular, if
\begin{enumerate}
\item the initial Wigner function is Gaussian, 
\item the Hamiltonian is at most quadratic with respect to the
  continuous-variable operators, such that the Heisenberg equations of
  motion for these operators are linear, and
\item the measurements can be modeled as von Neumann measurements of
  arbitrary linear combinations of the continuous variables,
\end{enumerate}
the quantum observation statistics become equivalent to those of an
HGMM described in Appendix~\ref{hgmm}
\cite{wiseman_milburn,braunstein_rmp,bartlett2012}, and all the
statistical methods valid for an HGMM are also applicable to such a
quantum model. Sec.~\ref{optomech} is an example of how this
correspondence can be exploited for the purpose of hypothesis testing
and parameter estimation.

\section{\label{open}Open quantum systems}
The concepts introduced in this section can also be found in many
textbooks \cite{breuer,nielsen,wiseman_milburn,gardiner_zoller}.

\subsection{Density operator}
We would now like to incorporate more probabilities to model classical
uncertainties in $\psi_0$.  Suppose that $\psi_0$ depends on a
classical hidden variable $j$, and the probability distribution for
$j$ is $P^j$. $P(\psi)$ becomes
\begin{align}
P(\psi) &= \sum_j P(\psi|\psi_{0}^j)P^j
\\
&= \sum_j \Avg{\psi,\psi_{0}^j}P^j
\\
&= \Avg{\psi,\rho_0},
\label{born_density}
\end{align}
where 
\begin{align}
\rho_0 &= \sum_j P^j \psi_{0}^j \equiv \sum_j P^j \ket{\psi_{0}^j}\bra{\psi_{0}^j}
\label{rho0}
\end{align}
is called a density operator.  

Another way of arriving at the density operator is to consider a
larger Hilbert space as a tensor product of two smaller
ones $A$ and $B$, with an initial state given by $\Psi_0$ and a
final von Neumman projection given by
\begin{align}
\Psi &\equiv \psi\otimes \psi_B,
\\
P(\Psi|\Psi_0) 
&= \Avg{\Psi,\Psi_0}_{AB} = \Avg{\psi\otimes\psi_B,\Psi_0}_{AB}.
\end{align}
If we neglect the outcome $\psi_B$, the marginal probability
function is
\begin{align}
P(\psi) &= \sum_{\psi_B} P(\Psi|\Psi_0) 
= \Avg{\psi \otimes I_B,\Psi_0}_{AB}
=\Avg{\psi,\rho_0}_A,
\end{align}
where $I_B$ denotes the identity operator in $B$, and
\begin{align}
\rho_0 &= \Avg{I_B,\Psi_0}_B \equiv \trace_B \ket{\Psi_0}\bra{\Psi_0}
\end{align}
turns out to have exactly the same properties as Eq.~(\ref{rho0}).

The third and the most nontrivial way of arriving at a density
operator is Gleason's theorem \cite{gleason}, which roughly states
that, if $N\ge 3$ and we are \emph{given} a probability function
$P(\psi)$, then there always exists a positive-semidefinite operator
$\rho_0$ such that Eq.~(\ref{born_density}) holds. The theorem is
redundant, however, if we assume Born's rule, as we have already
derived Eq.~(\ref{born_density}) by other more constructive means and
we do not really need the theorem to tell us that a $\rho_0$ exists.

It is common in quantum physics to call $\rho_0$ a quantum state; it
is called a pure state when $\rho_0 = \psi_0^j \equiv
\ket{\psi_0^j}\bra{\psi_0^j}$ is a projection and a mixed state
otherwise. This terminology is confusing and we shall avoid it here,
as the density operator is different from the state concept in
probability theory, as described in Appendix~\ref{state}.

\subsection{Positive operator-valued measure (POVM)}
The von Neumann projection can be generalized to a more general notion
of measurement called the POVM $E(y)$, where $y$ is an
observation. The POVM is a positive-semidefinite operator that
satisfies the completeness property:
\begin{align}
\sum_{y} E(y) &= I,
\end{align}
with $I$ denoting the identity operator. The probability function of
$y$ is then given by
\begin{align}
P(y) &= \Avg{E(y),\rho_0}.
\end{align}
It can be shown that a POVM is equivalent to a von Neumann projection
in a larger Hilbert space, but it is a convenient tool nonetheless to
model partial measurements.

\subsection{Time evolution}
Instead of the unitary map, we can use a more general mathematical
operation called a completely positive (CP) map to model dynamics that
involve uncertainties:
\begin{align}
P &= \Avg{E,\mathcal V\rho_0} = \Avg{\mathcal V^* E,\rho_0},
\end{align}
where the trace-preserving CP map $\mathcal V$ can be written in the
Kraus representation as
\begin{align}
\mathcal V \rho_0 &= \sum_j V_j \rho_0 V_j^\dagger.
\end{align}
$V_j$ is called a Kraus operator, which satisfies the completeness
property:
\begin{align}
\sum_j V_j^\dagger V_j &= I.
\end{align}

To model continuous-time evolution, a CP map can be written
as 
\begin{align}
\mathcal V &= \exp(\mathcal Lt),
\end{align}
where $\mathcal L$ is known as the Lindblad generator.

A CP map can be used to describe the phenomenon of decoherence, which
occurs when the system of interest interacts with another inaccessible
system. The system of interest is then called an open system.  Like
the density operator and the POVM, it can be shown that a CP map is
equivalent to unitary evolution in a larger Hilbert space that
includes all the inaccessible subsystems.

\subsection{Generalized measurements}
For a series of generalized measurements, the probability function of
the outcome can be written as
\begin{align}
P(Y) &= P(y_K,\dots,y_1)
\\
&=\Avg{E(y_K),\mathcal W(y_{K-1})\dots\mathcal W(y_1)\rho_0}
\label{joint}
\end{align}
where $\mathcal W(y_k)$ is a CP map with an observation $y_k$ at time
$k$, describing both the dynamics and the probabilities of the
observation. It reduces to a trace-preserving CP map $\mathcal V_k$
when summed over all possible outcomes:
\begin{align}
\sum_{y_k} \mathcal W(y_k) &= \mathcal V_k.
\end{align}
In principle, Eq.~(\ref{joint}) can also be modeled using
Eq.~(\ref{qm_nutshell}) in a larger Hilbert space through the
principle of deferred measurement, but for numerical analysis a
smaller Hilbert space is usually more desirable to alleviate the curse
of dimensionality. Eq.~(\ref{joint}) may be regarded as a
generalization of the classical HMM.

We have stressed repeatedly that the open quantum system theory is a
reformulation of quantum probability theory and contains no new
physics, but its value for fundamental physics should not be dismissed
entirely. After all, Hamiltonian and Lagrangian mechanics were also
merely reformulations of Newtonian mechanics, until quantum mechanics
turned them into a centerpiece.

\section{\label{qest}Quantum estimation}
\subsection{Quantum filtering}
The goal of quantum filtering is to predict the future observation
$y_{k+1}$ for \emph{any} given $E(y_{k+1})$ using the past
observations $Y_{k} \doteq \{y_{k},\dots,y_1\}$.  Using
Eq.~(\ref{joint}), the filtering probability function becomes
\begin{align}
&\quad P(y_{k+1}|Y_k) 
\nonumber\\
&= \frac{P(y_{k+1},Y_k)}{ P(Y_k)}
\\
&= \frac{\Avg{E(y_{k+1}),\mathcal W(y_{k})\dots\mathcal W(y_1)\rho_0}}
{\Avg{I,\mathcal W(y_{k})\dots\mathcal W(y_1)\rho_0}}
\\
&= \Avg{E(y_{k+1}),\rho(Y_k)},
\end{align}
where the posterior density operator defined as
\begin{align}
\rho(Y_k) &= \mathcal C\mathcal W(y_{k})\dots\mathcal W(y_1)\rho_0,
&
\mathcal C\rho \doteq \frac{\rho}{\Avg{I,\rho}},
\label{qbayes}
\end{align}
contains the sufficient statistics for filtering.  Eq.~(\ref{qbayes})
is sometimes called the quantum Bayes theorem
\cite{gardiner_zoller}. A useful way of computing Eq.~(\ref{qbayes})
is to find a classical HMM representation via quasiprobability
functions and take advantage of existing classical algorithms. The
Kalman filter is especially useful for quantum optomechanics and large
atomic spin ensembles \cite{wiseman_milburn}, as we have also seen
from Sec.~\ref{optomech}.

As pioneered by Belavkin \cite{belavkin}, a continuous-time limit of
Eq.~(\ref{qbayes}) can be defined using stochastic calculus to model
observations with white noise \cite{wiseman_milburn}.  See also
Ref.~\cite{bouten} for an alternative mathematical treatment of
continuous-time quantum filtering.

Quantum filtering can form the basis for quantum parameter estimation
and hypothesis testing techniques; see, for example,
Refs.~\cite{gambetta2001,hypothesis}.

\subsection{\label{smooth}Quantum smoothing}
Quantum smoothing is the estimation of $y_{k+1}$ using
the past
\begin{align}
Y_k &= \BK{y_k,\dots,y_1},
\end{align}
as well as the future
\begin{align}
\bar Y_{k+1} &= Y \setminus Y_{k+1} = \BK{y_K,\dots,y_{k+2}},
\end{align}
assuming that $y_{k+1}$ is missing. The conditional probability
function is
\begin{align}
P(y_{k+1}|Y_k,\bar Y_{k+1}) &=\mathcal N P(\bar Y_{k+1},y_{k+1},Y_k),
\end{align}
where $\mathcal N$ is a normalization constant. We rewrite
Eq.~(\ref{joint}) in the time-symmetric form in terms of
Eq.~(\ref{qbayes}):
\begin{align}
P(Y) &= \Avg{E(\bar Y_{k+1}),\mathcal W(y_{k+1})\rho(Y_k)},
\\
E(\bar Y_{k+1}) &= \mathcal W^*(y_{k+2})\dots \mathcal W^*(y_{K-1})E(y_K).
\label{E}
\end{align}
Eq.~(\ref{E}) has the same structure as the filtering equation in
Eq.~(\ref{qbayes}) and can be calculated by the same methods applied
backwards in time.  Hence
\begin{align}
P(y_{k+1}|Y_k,\bar Y_{k+1})
&= \mathcal N\Avg{E(\bar Y_{k+1}),\mathcal W(y_{k+1})\rho(Y_k)},
\label{tsym}
\end{align}
and $\rho(Y_k)$ and $E(\bar Y_{k+1})$ provide the sufficient
statistics for smoothing. Continuous-time limits of quantum smoothing
can be found in
Refs.~\cite{smooth,smooth_pra1,smooth_pra2,gammelmark2013}.

The omission of $y_{k+1}$ from the given observations may seem
artificial, but this formulation can actually be used for quantum
sensing of hidden classical waveforms. This is done by embedding a
classical HMM in the quantum model and assuming that $\mathcal
W(y_{k+1})$ is a perfect observation of the classical HMM
\cite{smooth,smooth_pra1,smooth_pra2}. Recent quantum optics
experiments that used smoothing for waveform estimation are reported
in Refs.~\cite{wheatley,yonezawa,iwasawa}.

The concept of quantum smoothing can be traced back to Aharonov
\textit{et al.}\ \cite{abl}, who proposed the time-symmetric form
given by Eq.~(\ref{tsym}) for von Neumann measurements. The connection
between this time-symmetric form and smoothing estimation was first
made and studied by Tsang \cite{smooth,smooth_pra1,smooth_pra2}.  The
presentation here follows a more recent work by Gammelmark \textit{et
  al.}\ \cite{gammelmark2013}.

The curse of dimensionality also exists for quantum estimation, as the
number of variables that specify a density matrix also grows
exponentially with the degree of freedom. As quantum technologies
become more complex and nonclassical, one can envision an increasing
demand for efficient quantum filtering and smoothing algorithms for
future signal processing and control applications.

% \subsection{Quantum metrology}
% A lot of the statistical techniques presented in this paper are also
% relevant to quantum metrology, which is the study of precision
% measurements \emph{assuming} that quantum mechanics is correct.
% Since the pioneering studies by Helstrom \cite{helstrom}, Holevo
% \cite{holevo}, Braginsky and Khalili \cite{braginsky}, Caves and
% Thorne \cite{caves}, and many others, significant theoretical
% \cite{wiseman_milburn,paris_rehacek,glm_science,glm2011,twc,stellar,qzzb,tsang_nair,tsang_open}
% and experimental progress
% \cite{ligo2011,schnabel,wheatley,yonezawa,iwasawa} has been made to
% generalize and apply the theory to more realistic systems, including
% optomechanics. It would be interesting to explore whether quantum
% metrology may also be relevant to tests of quantum mechanics itself:
% Refs.~\cite{downes,doukas} are relevant work in this direction.

%\section{Quantum hypothesis testing}
%\subsection{Hybrid filtering approach}

%\subsection{Likelihood ratio formula}

%\subsection{Relative entropy and mismatched mean-square error}

%\bibliography{research}
\bibliography{testing_quantum3}
%\printbibliography

\end{document}